\documentclass[preprint,doublespace,authoryear,3p,times,10pt]{elsarticle}

\usepackage{multirow,setspace,times,amssymb,amsmath,graphicx,color,rotating,subfigure,url}
\usepackage{lineno}
\usepackage{natbib}
\usepackage{rotating}
\usepackage{mathrsfs}
\usepackage{makecell,booktabs,longtable}
\usepackage[bookmarks=true,colorlinks,linkcolor=red,anchorcolor=blue,citecolor=green,unicode]{hyperref}
\usepackage{bookmark}

\usepackage{dcolumn}
\newcolumntype{d}[1]{D{.}{.}{#1}}

\journal{Journal of International Financial Markets, Institutions \& Money}

\begin{document}

\begin{frontmatter}

\title{Cross-shareholding networks and stock price synchronicity: Evidence from China}

\author[CSU,UW,UE]{Fenghua Wen}
\author[CSU]{Yujie Yuan}
\author[DF,DM]{Wei-Xing Zhou\corref{WXZ}}
\cortext[WXZ]{Corresponding author. 130 Meilong Road, P.O. Box 114, School of Business, East China University of Science and Technology, Shanghai 200237, China.}
\ead{wxzhou@ecust.edu.cn}

\address[CSU]{School of Business, Central South University, Changsha, Human 410083, China}
\address[UW]{Supply Chain and Logistics Optimization Research Centre, Faculty of Engineering, University of Windsor, Windsor, ON, Canada}
\address[UE]{Centre for Computational Finance and Economic Agents, University of Essex, Colchester CO4 3SQ, UK}
\address[DF]{Department of Finance, East China University of Science and Technology, Shanghai 200237, China}
\address[DM]{Department of Mathematics, East China University of Science and Technology, Shanghai 200237, China}

\begin{abstract}
This paper investigates the effect of cross-shareholding on stock price synchronicity, as a measure of price informativeness, of the listed firms in the Chinese stock market. We gauge firms' levels of cross-shareholdings in terms of centrality in the cross-shareholding network. It is confirmed that it is through a noise-reducing process that cross-shareholding promotes price synchronicity and reduces price delay. More importantly, this effect on price informativeness is pronounced for large firms and in the periods of market downturns. Overall, our analyses provide insights into the relation between the ownership structure and price informativeness.
\end{abstract}


\end{frontmatter}


\section{Introduction}
\label{S1:Introduction}

With increasing economic integration, firms relying only on their own capital accumulation to achieve the goal of profit and expansion have gradually lost the advantage to pull ahead in the long run. Given the limited resources available to individual firms and the complexity of financial markets, firms often hold part of other firms' shares and allow themselves to be partly owned in order to utilize more resources and further foster strategic alliances to obtain more benefits \citep{Amundsen-Bergman-2002-EJ}. Many simple and theoretical estimations of firms' equity values under this special ownership structure has been made \citep{Elsinger-2011,Fischer-Tom-2013-MF}. Meanwhile, when focusing on corporate efficiency, many studies suggest that cross-shareholding would have an impact on the information environment. For example, there exists a stricter monitoring system among firms with cross-shareholdings \citep{Sheard-1994-JFSCS,Jiang-Kim-2000-IJA} and such cross-shareholding can reduce the information asymmetry among shareholders \citep{Brooks-Chen-Zeng-2018-JCF,Liu-Lin-Qin-2018-EL}. Furthermore, such an effect is attributable to a process of reducing moral hazard and restraining shareholders' opportunistic behavior \cite{Ang-Constand-Mathur-Booth-2002-JMFM}. As mentioned in previous studies, the improvement of the information environment as a result of firms' characteristics (e.g., analyst coverage and ownership concentration) can incorporate more firm-specific information into stock prices \citep{Jin-Myers-2006-JFE,Gul-Kim-Qiu-2010-JFE}, and may drive irrational factors out \citep{Lee-Liu-2011-JBF}, further impacting on the price informativeness. However, evidence of the impact of the information environment (arising from cross-shareholding) on the price informativeness is rare in academic research.

In terms of stock price informativeness, the existing literature employs ${R^2}$ to measure either firm-specific, or market-wide information in the stock prices \citep{Roll-1988-JF,Morck-Yeung-Yu-2000-JFE,Jin-Myers-2006-JFE} and finds that firms in emerging economies, where it is confirmed to have poor protection of investors and institutional environment, have a higher ${R^2}$ than those in the developed economies. Furthermore, a growing body of studies have investigated how firms' characteristics, or their institutional environment affect individual stock price informativeness \citep{Hutton-Marcus-Tehranian-2009-JFE,Gul-Kim-Qiu-2010-JFE,Dong-Li-Lin-Ni-2016-JFQA}. The rationale for these researches is that a more transparent information environment results in more firm-specific information being incorporated into stock prices such that we would observe a lower level of stock price synchronicity. However, this information-based perspective, while intuitive, is at odds with other implications of market efficiency. Some studies also argue that stock return synchronicity is positively related to the information environment due to analyst coverage and noise trading \citep{Chan-Hameed-2006-JFE,Lee-Liu-2011-JBF,Crawford-Jones-Roulstone-So-2012-AR}. In this way, if the improvement of the information environment arising from cross-shareholding accelerates the production of firm-specific information, we might observe a negative relation between cross-shareholding and stock price informativeness. If the improvement of the information environment drives the irrational factors out, we expect that the stock price informativeness will increase with firms' centrality in the cross-shareholding network. It is necessary for us to reveal how the cross-shareholding affects the price informativeness.

In this paper, we utilize a unique set of data that documents and introduces cross-shareholding of Chinese stocks. The documents (which were downloaded from a database) contain the listed firms and their shareholders. Cross-shareholding is treated as a combination of listed firms and their shareholders and cross-shareholders are defined as two firms that are shareholders of each other. Therefore, there exists a network filled with equity ties in the stock market. The nodes, representing individual firms, and the linkages, representing the cross-shareholding between firms, form a cross-shareholding network. In a network framework, researchers find various relations between the financial linkages and individual performance, such as firm performance \citep{Cai-Sevilir-2010-JFE,Larcker-So-Wang-2013-JAE,Rossi-Blake-Timmermann-Tonks-Wermers-2018-JFE} and compensation for risk taking \citep{Ahern-2013,Aobdia-Caskey-Ozel-2014-RAS,Herskovic-2018-JF}. Meanwhile, the networks based on ownership linkages have attracted much attention from researchers. For example, \cite{Khanna-Thomas-2009-JFE} find that the pairwise stock price synchronicity is strong between firms with equity ties. Using complex network theory, \cite{Ma-Zhuang-Li-2011-PA} initially measure the cross-shareholding of listed companies in the Chinese stock market. It is further unveiled that the cross-shareholding network can reveal complex relationships precisely in the stock market and the network exhibits small-world and scale-free properties \citep{Li-An-Gao-Huang-Xu-2014-PA,Chang-Wang-2017-DDNS}. Therefore, following previous studies, we can apply complex network theory to gauge the level of cross-shareholding to further reveal the impact of cross-shareholding on the price informativeness.

Due to the rapid growth and increasing reputation of the Chinese financial market among the global financial markets, scholars and investors have paid much attention to the Chinese market. It is also widely acknowledged that the Chinese stock market has several unique characteristics that challenge the traditional asset pricing theories. For example, the trading activities of small retail investors have generally played a leading role in the stock market and the institutional traders accounted for less than 20\% of the total trading volume at the end of 2016. In terms of the cross-shareholding, while it is widely observed in developed economies and is a crucial characteristic of some large business groups, more listed firms in mainland China have been permitted to hold shares of other firms after the split-share reform in 2005. Moreover, most of the firms in the Chinese stock market mutually holding shares aim to have better capital operations or superior opportunities of the initial public offering. Some of them also act as short-term arbitragers through these non-principal investments---especially when the stock market experiences a boom---while a small amount of them can find a long-term investing relationship.

Motivated by these pioneering researches on cross-shareholding and price informativeness, the purpose of this paper is to examine, from the network perspective, the impact of cross-shareholding on stock price informativeness, measured as stock price synchronicity. First, we carry out a univariate analysis as well as a multivariate analysis to illustrate the point that the more transparent the information environment of a firm as a result of its more central position in the cross-shareholding network, the more synchronous its stock price. This is because more fundamental information is already available to investors in a better information environment, such that the stock price contains fewer noises and pricing errors. Moreover, a more transparent information environment can help investors improve their predictions about the occurrence of future firm-specific events, therefore reducing the likelihood of unexpected news \citep{Dasgupta-Gan-Gao-2010-JFQA}. To sharpen the inferences of our main empirical tests, we examine alternative measures of stock price synchronicity and our results remain unaffected.

Second, while we can conclude that cross-shareholding has a positive impact on the stock price synchronicity, it is also skeptical that whether cross-shareholding would really improve stock price informativeness since firms with higher levels of cross-shareholding can be affected by the operating conditions of other firms. Put another way, instead of reducing irrational factors, the stocks of firms with a more central position in the cross-shareholding network reflect more fundamental information, which, ceteris paribus, promotes the stock price synchronicity. To ensure that the observed positive impact on the stock price synchronicity is driven by a reduction in irrational factors, we perform a complementary analysis. We find a negative relation between cross-shareholding and noise trading, corroborating the proposition that there exists a noise-reducing process in the impact of the information environment arising from cross-shareholding on the price informativeness in the Chinese stock market. Our conclusions still hold after we control for the weekly returns of the three factors of \cite{Fama-French-1993-JFE} as the fundamental values.

Third, based on the confirmation of the noise-reducing process in stock price informativeness, we further examine the impact of cross-shareholding on price delay, measured as the proportion of the previous market-wide information reflected in the current price \citep{Hou-Moskowitz-2005-RFS} and also treated as an alternative measure of the stock price informativeness \citep{Dong-Li-Lin-Ni-2016-JFQA}. Researchers argue that such a delay can be caused by a potential market friction and pricing inefficiency. Following their approach, we find that firms---especially large firms---with a more central position in the cross-shareholding network have a lower level of price delay. These results indicate that firms with a more central position in the cross-shareholding network would have a better-quality information environment and lower levels of noise trading. Hence, their stock prices would thus contain more contemporary information instead of historical information and the stocks would be more rationally priced. Furthermore, we believe that investors might pay attention to firms' cross-shareholders through publicly available data (e.g., the balance sheets). Consequently, based on the improvement of the information environment as a result of  cross-shareholding for large firms, the increasing investor attention from other firms may strengthen the impact of the reduction in noise trading, and thus firms with more cross-shareholding would have a lower level of price delay. These results add further supports for our main inference that cross-shareholding reduces noise trading.

Fourth, we find that such an positive effect on the stock pricing efficiency is pronounced for large firms. This is because the stocks of large firms that hold a more central position in the cross-shareholding network tend to be priced more rationally and therefore reflect more about contemporaneous market-wide information, while more irrational factors are clustered at small firms \citep{Zhang-2006-JF,Zhang-Cai-Keasey-2013-JAccE}. Additionally, a feasible interpretation of our results is the idea that, compared to the objectives of small firms when they hold other firms' stock (e.g., short-term financing and arbitrage), large firms have more stable future cash flows and are less sensitive to market-wide shocks. Therefore, investors prefer to invest in a long-term horizon, which strengthens the alignment effect between the shareholders and demands a higher-quality information environment.

Last but not least, based on the Adaptive Markets Hypothesis that market efficiency varies from time to time \citep{Lo-2004-JPM}, we find that, while cross-shareholding has a small effect on stock price informativeness during market upturns, the effect becomes stronger during market downturns. Specifically, investing behavior in a financial market is a type of dynamic decision-making behavior \citep{Wen-Gong-Chao-Chen-2014-MPE} and it is argued that there exists a persistent overreaction during market downturns and a persistent underreaction during market upturns \citep{Hou-Xiong-Peng-2009}. A feasible interpretation of our results is the idea that investors in the Chinese market would pay more attention to the portion of firm information related to market-wide information during a market downturn, which is consistent with the bargain-hunting or stop-loss strategies. Therefore, investors accelerate the incorporation of such market-wide information into the stock prices, and thus strengthen the impact of the reduction in noise trading on the price informativeness during market downturns. Our results remain unaffected when using other centrality measures and when considering the potential selection bias caused by short selling.

Our findings contribute to the extant literature in three ways. First, while previous studies focus on the impact of cross-shareholding on corporate management and decision making, we provide intuitive and fresh evidence that cross-shareholding has a further impact on the variation of stock price through the improvement of the information environment and a further reduction in noise trading. Moreover, our results add to the theoretical prediction in the \cite{Lee-Liu-2011-JBF} model that the information environment positively affects stock price synchronicity. Second, we add complementary results to the impact of firms' ownership structure on stock price's behavior, even if there is a growing body of empirical studies investigating the impact of ownership structure, such as director linkage \citep{Khanna-Thomas-2009-JFE}, ownership concentration \citep{Gul-Kim-Qiu-2010-JFE}, block ownership \citep{Brockman-Yan-2009-JBF}, and large controlling shareholders \citep{Boubaker-Mansali-Rjiba-2014-JBF}, on stock price informativeness. Third, apart from enriching the research on financial markets using a network framework \citep{Xi-An-2018-PA,Herskovic-2018-JF}, our study is one of the few---if not the first---to employ complex network theory to investigate the relation between corporate structure and stock pricing efficiency\footnote{Others focus on the networks of boardrooms, underwriters, chief executive officers (CEO), and other social relations \citep{Cohen-Frazzini-Malloy-2008-JPE,Jiang-Zhou-2010-PA,Larcker-So-Wang-2013-JAE,Ozsoylev-Walden-Yavuz-Bildik-2014-RFS,El-Fogel-Jandik-2015-JFE,Bajo-Chemmanur-Simonyan-Tehranian-2016-JFE,Adamic-Brunetti-Harris-Kirilenko-2017-EmJ,Cheng-Felix-Zhao-2019-JBF}.}.

The rest of the paper is structured as follows. We relate our work to existing literature in Section \ref{S1:LitRev}. Section \ref{S1:Data} describes the setting up of the cross-shareholding network, the data sources, and the construction of variables. Section \ref{S1:Empirical} reports our empirical results, which are further discussed in Section \ref{S1:Discuss}. We offer our conclusions in Section \ref{S1:Conclude}.

\section{Literature review}
\label{S1:LitRev}

This study sits at the intersection of two important literature streams. The first stream investigates the link between information efficiency and stock price informativeness. An innovation in this field was suggested by \cite{Roll-1988-JF} that the fundamental value is determined by both market-wide information and firm-specific information, where stock price synchronicity is treated as a measure of stock price informativeness. This groundbreaking idea has motivated several follow-up studies that examine stock price informativeness around the world. \cite{Morck-Yeung-Yu-2000-JFE} find that China has the second-highest ${R^2}$ in the world, and that it is the relatively poor protection of investors' property rights that make the stock price synchronicity (${R^2}$) in emerging markets higher than that in developed markets. By replicating the results of \cite{Morck-Yeung-Yu-2000-JFE} and presenting several vivid stories, \cite{Jin-Myers-2006-JFE} argue that in certain improbable but noteworthy circumstance, the impact of investor protection is doubtful and firms' opaqueness does make a difference. \cite{He-Li-Shen-Zhang-2013-JIMF} find that the positive impact of large foreign ownership on price informativeness may be stronger in developed economies with strong investor protection and a transparent information environment. In addition, \cite{Dang-Moshirian-Zhang-2015-JFE} measure the commonality of market-wide information and firm-specific information around the world.

Extending the cross-country pattern of stock price informativeness to the firm level, \cite{Hutton-Marcus-Tehranian-2009-JFE} find a result akin to that of \cite{Jin-Myers-2006-JFE} by giving a detailed measure of firms' opaqueness in the accounting item. \cite{Piotroski-Roulstone-2004-AR} make the first attempt to investigate the relation between analyst coverage and stock price synchronicity and \cite{Liu-2011-JFQA} and \cite{Crawford-Jones-Roulstone-So-2012-AR} find that analyst coverage accelerates the incorporation of firm-specific information into stock prices. In terms of the ownership structure, ownership concentration and large controlling shareholders are proven impediments to the incorporation of firm-specific information into stock prices \citep{Gul-Kim-Qiu-2010-JFE,Boubaker-Mansali-Rjiba-2014-JBF}. \cite{Ben-Cosset-2014-JCF} further posit that state ownership results in a less transparent environment, making the collection of private firm-specific information costly and discouraging informed trading. Other studies investigate the impact of seasoned equity offerings (SOEs), institutional investors, corporate social responsibility and independent director reputation incentives on stock price synchronicity \citep{An-Zhang-2013-JCF,Becchetti-Ciciretti-Hasan-2013-JCF,Bai-Hu-Liu-Zhu-2016-JFS,Sila-Gonzalez-Hagendorff-2017-JCF}. These researches have the common perspective that a more transparent information environment produces more firm-specific information, weakens the explanatory power of market-wide information, and thus reduces stock price synchronicity.

Turning to the discussion of whether stock co-movements are explained by firm-specific information or noise, contrary to the aforementioned views, small and young firms, which are infrequently traded, are followed by a low level of stock price synchronicity \citep{Kelly-2007}. \cite{Chan-Hameed-2006-JFE} and \cite{Crawford-Jones-Roulstone-So-2012-AR} argue that it is the analyst coverage in emerging economies that promotes the production of the market-wide information, which improves pricing efficiency. In addition, a more transparent information environment can help investors improve their predictions about the occurrence of future firm-specific events, and therefore reducing the likelihood of unexpected news, which further promotes stock price synchronicity \citep{Dasgupta-Gan-Gao-2010-JFQA}.

Our study is also related to the paper that documents a U-shaped relation between price informativeness and stock price synchronicity \citep{Lee-Liu-2011-JBF}. Through a multi-asset, multi-period noisy rational expectations equilibrium, \cite{Lee-Liu-2011-JBF} find that noise trading reduces stock price informativeness.
In this paper, we provide empirical evidence supporting their theoretical prediction of the impact of noise trading\footnote{More substantively, according to \cite{Lee-Liu-2011-JBF}, there exists a U-shaped relation between return volatility and price informativeness in a market where informed investors lead stock trading, while return volatility has a strictly monotonic decreasing association with price informativeness in a market filled with noise traders.}.

The second stream of literature is related to cross-shareholding networks. Cross-shareholding has impacts on decision making. For example, \cite{Osano-1996-JBF} finds that cross-shareholding reduce the problem of shortsightedness caused by outside takeovers. \cite{Ang-Constand-Mathur-Booth-2002-JMFM} suggest that long-term trading relations established through cross-shareholding largely reduce moral hazard among shareholders, and restrain the opportunistic behavior in the trading contract. Moreover, cross-shareholding not only changes shareholders' preferences but also is used as a strategy device to deter other firms' entries \citep{Harford-Jenter-Li-2011-JFE,Li-Ma-Zeng-2015-EL}.

In terms of the information environment, firms with more cross-shareholding can utilize more information and financial resources, and there exists a stricter monitoring system among them, which further reduces the information asymmetry between shareholders \citep{Sheard-1994-JFSCS}. \cite{Jiang-Kim-2000-IJA} conjecture that firms with higher levels of cross-shareholding have lowers level of information asymmetry between corporation and market participants, which makes the stock price incorporate information about future profitability much faster. As a for of indirect cross-shareholding between firms caused by institution cross-shareholding, this type of ownership structure not only reduces deal premiums, thus leading to better value for acquirers, but also provides superior two-sided information and better monitoring roles \citep{Brooks-Chen-Zeng-2018-JCF}. Moreover, for both high-cost and low-cost firms, it is the shares held in each other that provide two-way information communication and incentives to reveal private information, which improves firms' information disclosure and reduces information asymmetry \citep{Liu-Lin-Qin-2018-EL}. Hence, according to the aforementioned literature, the primary goal of this research is to investigate the relation between cross-shareholding and price dynamics through the impact of cross-shareholding on the information environment, while the literature has given plenty of theoretical evidence of the impact of cross-shareholding on the estimation of equity values \citep{Elsinger-2011,Fischer-Tom-2013-MF}.

With the development of comprehensive cross-business among financial institutions and firms, networks filled with financial linkages have become an indispensable factor when studying individual performance and price dynamics \citep{Li-Jiang-Xie-Xiong-Zhang-Zhou-2015-PA}. A related paper constructs a correlation network in the stock market and focuses on the static topological properties of the network, while neglecting the analysis of the potential consequences \citep{Xi-An-2018-PA}.
Based on the financial relations, some papers further investigate how firms or portfolios perform from a network perspective \citep{Hochberg-Ljungqvist-Lu-2007-JF,Cai-Sevilir-2010-JFE,Rossi-Blake-Timmermann-Tonks-Wermers-2018-JFE}. For example, firms with more director linkages tend to collect information more efficiently and demand less from the consultation services of investment banks, which promotes profits even after the conclusion of merger and acquisition activities \citep{Larcker-So-Wang-2013-JAE}.

Given that individuals in financial networks can use more financial resources and update their strategies in order to optimize their investment profits, some works reveal the impact of their trading behavior on stock pricing and individual risk bearing. For example, \cite{Ahern-2013} and  \cite{Aobdia-Caskey-Ozel-2014-RAS} conjecture that, due to the adoption of more fundamental information, firms or sectors at more central positions in a business-related network would have higher compensation for the systematic risk. Additionally, considering an input-output network, \cite{Herskovic-2018-JF} further finds that changes in the network are sources of systematic risk and that taking sparsity and concentration of the network into account can help fill a significant return gap that cannot be explained by standard asset pricing models such as the capital asset pricing model (CAPM) or the \cite{Fama-French-1993-JFE} three-factor model. These papers believe that, compared with individuals isolated from the outside world, it is through financial networks where individuals communicate with each other and exchange the information that the future expectation and equilibrium market prices would be changed.

    Our study is also related to several researches that measure the activities of cross-shareholding in a network framework. For example, \cite{Ma-Zhuang-Li-2011-PA} make the first attempt to establish a cross-shareholding network of listed companies in the Chinese stock market and shows that the cross-shareholding network has scale-free characteristics at both the firm level and the province level. Finding a similar pattern to the results of the former study, \cite{Li-An-Gao-Huang-Xu-2014-PA} and \cite{Chang-Wang-2017-DDNS} suggest that cross-shareholding networks reveal complex relationships precisely in the stock market and further find a law of evolution of the cross-shareholding network structure. However, these researches focus either on the establishment of networks or the topological characteristics, but overlook the analysis of further economic consequences. In contrast, our study tries to measure the level of cross-shareholding in a broad sense to further reveal the impact of cross-shareholding on the price informativeness.

\section{Data and methodology}
\label{S1:Data}

\subsection{Measure of stock price synchronicity}

Following \cite{Hutton-Marcus-Tehranian-2009-JFE}, an expanded index model is applied to measure stock price synchronicity, which allows us to decompose the total return variations into two components: those tied to common (market wide and/or industry wide) factors and those tied to unobserved factors. We first estimate the following model for each fiscal year:
\begin{equation}
{r_{j,t}} = {\alpha _j} + {\beta _{1,j}}{r_{m,t - 1}} + {\beta _{2,j}}{r_{i,t - 1}} + {\beta _{3,j}}{r_{m,t}} + {\beta _{4,j}}{r_{i,t}} + {\beta _{5,j}}{r_{m,t + 1}} + {\beta _{6,j}}{r_{i,t + 1}} + {{\rm{\varepsilon }}_{j,t}}
\label{Eq:RegEq}
\end{equation}
where ${r_{j,t}}$ denotes the weekly return of firm \emph{j} at week \emph{t}, ${r_{m,t}}$ and ${r_{i,t}}$ denote respectively the value-weighted \emph{A+B} share market return and industry return,  and $\varepsilon$ represents the unspecified random factors. This \emph{A+B} share market return is based on the composite (value-weighted) \emph{A+B} share index which reflects \emph{A+B} share price movements in both the Shanghai and Shenzhen exchanges. The industry return is created by using all firms within the same industry. We allow for non-synchronous trading by including lead and lag terms for the market and industry indexes to alleviate concerns over potential non-synchronous trading biases \citep{Hutton-Marcus-Tehranian-2009-JFE,Gul-Kim-Qiu-2010-JFE}.
Since the ${R^2}$ is highly skewed and bounded between one and zero, we apply a logistic transformation to obtain a near normally distributed and continuous variable:
\begin{equation}
 {\emph{SYN}} = \ln \left(\frac{{{R^2}}}{{1 - {R^2}}}\right)
 \label{Eq:Syn}
\end{equation}
This \emph{SYN}-based efficiency measure has been used in various empirical studies on corporate investments and emerging markets \citep{Durnev-Morck-Yeung-2004-JF,Cheng-Leung-Yu-2014-IREF}. A higher value of \emph{SYN} indicates that the stock price is more synchronized.

\subsection{Construction of the cross-shareholding network and centrality measures}

Following \cite{Ma-Zhuang-Li-2011-PA} and \cite{Li-An-Gao-Huang-Xu-2014-PA}, we construct an undirected and unweighted cross-shareholding network of firms' direct cross-shareholders using cross-shareholding data from 2004 to 2016. Two companies are linked if and only if they are shareholders of each other. A firm's level of centrality is highly dependent on its position in the network. We focus on three basic measures of node centrality.

First, a firm is well-connected if it has many cross-shareholders. Such firms have more opportunities, greater access, or less information asymmetry than other firms. Degree centrality enumerates the number of first-degree links among cross-shareholders. Letting $\delta \left( {i,j} \right)$ denote an indicator that equals one when firms \emph{i} and \emph{j} are cross-shareholders and zero when their stocks are not mutually held and \emph{N} denote the total number of firms in the network over the course of a year, the degree centrality is obtained as follows
\begin{equation}
 Degree_i \equiv \frac{1}{N} \times \mathop \sum \limits_{j \ne i} \delta \left( {i,j} \right).
\end{equation}

Second, a firm may be well-connected if it lies on relatively more paths between any other pairs of outside firms, making such a firm crucial for connecting firms to each other. Betweenness centrality represents how important a firm is in connecting other firms to each other, or how well-situated a firm is in terms of the network paths it lies on \citep{Freeman-1977-Sociometry}. Betweenness centrality is defined as the average proportion of paths between two outside firms on which a firm lies. Let $\mathop P\nolimits_i (k,j)$ denotes the total number of shortest paths through firm \emph{i} between firm \emph{k} and firm \emph{j}, and $P(k,j)$ denotes the total number of shortest paths between \emph{k} and \emph{j}. The betweenness centrality is defined as follows:
\begin{equation}
 Between_i \equiv \frac{2}{{\left( {N - 1} \right)\left( {N - 2} \right)}} \times \mathop \sum \limits_{k \ne j:i \notin \left\{ {k,j} \right\}} \frac{{{P_i}\left( {k,j} \right)}}{{P\left( {k,j} \right)}}
\end{equation}
There is no difficulty to infer that firms with higher levels of degree centrality are more likely to have higher levels of betweenness centrality.

Finally, we consider a related notion of connectedness, stemming from \cite{Bonacich-1972-JMS}, as a refinement of degree centrality. A firm having more direct connections is more influential when the direct connections can reach or influence more outside firms. In other words, a firm is well connected when its direct contacts are also well connected. Eigenvector centrality measures a firm's connectedness based on the connectedness of its direct links:
\begin{equation}
 Eigen_i \equiv \frac{1}{\lambda}\mathop \sum \limits_j {g_{ij}} \times Eigen_j,
 \label{Eq:Eigen}
\end{equation}
where ${\rm{\lambda }}$ is the proportionality factor and $\mathop g\nolimits_{ij}  = 1$ if firms \emph{i} and \emph{j} are linked. Writing Eq.~(\ref{Eq:Eigen}) in vector form,
\begin{equation}
 {\rm{\lambda }} \times {\rm{Eigenvector}} = {\rm{G}} \times {\rm{Eigenvector}},
\end{equation}
we see that each firm's connectedness in the network can be obtained by the eigenvector of the matrix \emph{G}, where $G = \mathop {\{ \mathop g\nolimits_{ij} \} }\nolimits_{1 \le i,j \le N} $.
Since the location of a firm in the cross-shareholding network depends on the connectedness of its direct links, eigenvector centrality can be interpreted as capturing the notions of power and prestige. In other words, a firm may be well-connected when it is perceived to be prestigious and powerful, as it is more likely to be influenced by well-governed firms.

Although these three centrality measures can describe a firm's position in the network, one potential concern of our centrality measures is that larger firms with adequate resources and social relations tend to have higher levels of cross-shareholding. To isolate the centrality measures of the cross-shareholding network from firm size \citep{Larcker-So-Wang-2013-JAE}, we purge the three centrality measures of the effect of size by taking the residual from the cross-sectional regressions of \emph{Degree}, \emph{Eigen} and \emph{Between} on the log of firm size (\emph{Size}) and the square of log size, noting them as \emph{Rd}, \emph{Re} and \emph{Rb}, respectively.

\subsection{Data descriptive}

Considering that there has been an increasing number of cross-shareholding among listed firms since the split-shares reform was carried out in 2005, our sample period covers the thirteen-year period during 2004--2016. Among all the companies listed on either the Shanghai or the Shenzhen A-share stock exchanges, we collect the yearly cross-shareholding data from Wind Database by looking for the documents we are interested in, which include the list of listed company\footnote{Although the listed company list contains firms appearing on either the Shanghai or the Shenzhen A-share stock exchanges, some of their stockholders are listed on either the Shanghai or the Shenzhen B-share stock exchanges. Therefore, we employ the composite (value-weighted) A+B share index as a benchmark index to sufficiently cover market-wide information.},  the Main Stockholders List, and the Stockholder Relationships List.  Weekly stock returns and accounting data are obtained from the RESSET Financial Database (http://www.resset.cn/) and the China Stock Market and Accounting Research (CSMAR) Database (http://www.gtarsc.com/). Therefore, weekly stock returns can be assigned to each firm's fiscal year so as to match the time period of its reported financial data. We use weekly returns to avoid the problem of thinly traded stocks. We begin with all firm-years between 2004 and 2016 (11266 firm-years). Following previous researches \citep{Hutton-Marcus-Tehranian-2009-JFE,Gul-Kim-Qiu-2010-JFE,Bai-Hu-Liu-Zhu-2016-JFS}, we exclude firms marked with special treatment by the China Securities Regulatory Commission (CSRC) (259 firm-years), financial services firms and utilities (479 firm-years), firms' fiscal year fewer than 30 weeks of stock-return data (745 firm-years), firm-years with insufficient stock return data to calculate (1 firm-years), and firms with unavailable data of selected control variables (2495 firm-years). We are left with a final sample of 7287 firm-year observations. The sample covers 17 out of the 19 industries defined by the CSRC. The sample firm-years approximately increase across our sample period. See Table~\ref{TB:data} for details.

\begin{table}[h]
\centering
\caption{Sample distribution. Panel A shows the summary statistics of the cross-shareholding network of our initial sample. Diameters is the longest number of steps separating any two firms in the network. Path lengths is the distance measured by the smallest number of steps between any two firms. Clustering coefficient describes the connnectedness among a firm's cross-shareholders. Panel B shows the distribution of sample firms across industries based on the ``guidance on the industry category of listed companies' issued by the China Securities Regulatory Commission (CSRC), where A=Agriculture, B=Mining, C=Manufacturing, D=Electricity, gas, and water, E=Building and construction, F=Transportation and logistics, G=Information technology, H=Commerce, I=Real estate, J=Service, K=Culture and media, L=Conglomerate, M=Science research and technology service, N=Water conservancy,environment and public facilities management, P=Education, Q=Health and social work, R=Cultural, sports and entertainment, S=Comprehensive. Panel C shows the distribution by year.}
\medskip
\begin{tabular}{cccccccccccccc}
\hline
\multicolumn{8}{l}{Panel A: Summary statistics of the cross-shareholding network characteristics}\\
\hline
Year &2004& 2005& 2006& 2007& 2008& 2009& 2010\\
Firms &697 &700  &718  &817  &827  &887  &1004 \\
Links &1193&1185 &1245 &1581 &1461 &1877 &1890 \\
Diameters &13&14&14&11&12&14&12\\
Ave. path length &4.790&5.038&4.727&4.199&4.238&4.138 &4.528\\
Clustering	coef. &0.103&0.091&0.107&0.094&0.069&0.068 &0.051\\
                  &2011& 2012& 2013& 2014&2015& 2016\\
                  &976 &960  &994  &1025 &902 &905\\
                  &1726&1671 &1622 &1634 &1331&1267\\
                  &13&13&16&13&12&15\\
                  &4.484&4.556&4.715&4.660&4.860&5.238\\
                  &0.046&0.064&0.057&0.059&0.034&0.031\\
\hline
\multicolumn{8}{l}{Panel B: Industry distribution}\\
\hline
    & A & B & C & D & E & F& G & H  & I \\
\hline
$\#$ & 53  &267 & 3888 & 429   & 212& 659 & 427& 40&345 \\
$\%$ &0.73 &3.66 & 53.36&5.89  &2.91& 9.04 &5.86& 0.55&4.73 \\
& K & L & M & N & P & Q & R & S & Total\\
& 529&118&33&57&1&16&126&87&7287 \\
& 7.26&1.62&0.45&0.78&0.01&0.22&1.73&1.19&100 \\
\hline

\multicolumn{8}{l}{Panel C: Yearly distribution}\\
\hline
     &2004& 2005 &2006&2007 &2008& 2009&2010 \\
\hline
$\#$ & 223  &290 & 337 & 437   & 488& 527 & 651\\
$\%$ &3.06 &3.98 & 4.62&6.00  &6.70& 7.23 & 8.93 \\
\hline
     &2011 &2012&2013 &2014& 2015&2016 &Total\\
\hline
$\#$ &722& 768&825&789&597&633&7287 \\
$\%$ &9.91& 10.54&11.32&10.83&8.19&8.69&100 \\
\hline
\end{tabular}
\label{TB:data}
\end{table}

Panel A of Table~\ref{TB:data} shows the summary statistics of the cross-shareholding network characteristics of our initial sample. More than 600 firms have cross-shareholding each year with an average level of 1514 cross-shareholding across our sample. The average of diameters and the path lengths are 13.2308 and 4.6285, respectively, indicating that any two firms would be linked within an average of five other firms. We also find the average clustering coefficient to be 0.0672, indicating that around 7$\%$ of the time two firms linked by the same firm are also linked with each other.
Panel B of Table~\ref{TB:data} shows the distribution of our sample firms across industries based on the industry classification. Over 50$\%$ of the sample firms are from the manufacturing sector, followed by 9.04$\%$ from the transportation and logistics sector and 7.26$\%$ from the culture and media sector. Firms in the education industry account for 1$\%$ of the total.
Panel C reports the distribution of our sample firms by year. The number of firms increases over the early stage of the sample period, and the activities of cross-shareholding are affected by the crash of the Chinese stock market in 2015.

\subsection{Variables' descriptive statistics}

The data sample contains 7287 firm-years in the period during 2004-2016. For each firm-year, we regress Eq.~(\ref{Eq:RegEq}) and obtain its $R^2$. The corresponding {\emph{SYN}} is calculated according to Eq.~(\ref{Eq:Syn}). There are 7287 $R^2$ and {\emph{SYN}} in total. For each year during 2004-2016, we construct a cross-shareholding network and calculate the three centrality measures \emph{Degree}, \emph{Eigen} and \emph{Between}. The skewness (\emph{Skew}) and kurtosis (\emph{Kurt}) are calculated from the firm-specific weekly returns over each fiscal year. We also collect the following variables for each firm-year from 2004 to 2016: the log of size at the end of the previous fiscal year (\emph{Size}, the book-to-market ratio at the end of the previous fiscal year (\emph{Bm}), the leverage at the end of the previous fiscal year (\emph{Lev}), and the contemporaneous return on equity (\emph{Roe}).

Table~\ref{TB:VariableStat} presents the descriptive statistics for the key variables of interest. The mean and median ${R^2}$ are 0.5415 and 0.5498, respectively. The mean and median \emph{SYN} are 0.1821 and 0.2000, respectively. The \emph{SYN} measure is computed using the same specification of the market model introduced by \cite{Hutton-Marcus-Tehranian-2009-JFE}. The mean and the standard deviation of contemporaneous return on equity (\emph{Roe}) are respectively 6.9378 and 129.1204, indicating a significantly high variation in \emph{Roe} across the sample. The values of \emph{Degree}, \emph{Eigen}, and \emph{Between}, describing firms' positions in the network, exhibit a sharp increase from the median to the 95th percentile of the sample, which shows a fat-tailed phenomenon of the cross-shareholding network \citep{Ma-Zhuang-Li-2011-PA}. As we can see, in the cross-shareholding network, there exists a core-periphery structure: a small amount of firms are highly linked while most of the others have a few linkages.

\begin{table}[h]
\centering
 \caption{Descriptive statistics of the key variables of interest including ${R^2}$ , stock price synchronicity (\emph{SYN}), the log of size at the end of the previous fiscal year (\emph{Size}), the book-to-market ratio at the end of the previous fiscal year (\emph{Bm}), the leverage at the end of the previous fiscal year (\emph{Lev}), the contemporaneous return on equity (\emph{Roe}), the skewness (\emph{Skew}) and kurtosis (\emph{Kurt}) of the firm-specific weekly returns in each fiscal year, and the three cross-shareholding network centrality measures.}
 \medskip
 \begin{tabular}{lccccccc}
 \toprule
 & Mean  & Std. dev &5th Pctl.&25th Pctl.&Median&75th Pctl.&95th Pctl. \\
\hline
$\mathop R\nolimits^2 $  &0.5415&0.1625& 0.2580&0.4270 & 0.5498  &0.6634&  0.7948 \\
{\emph{SYN}}&  0.1821 &  0.7330& -1.0564 & -0.2939 &  0.2000 &  0.6785 &1.3542\\
{\emph{Skew}} &   0.6869 &   1.3366  &  -0.6481   & 0.0069 &0.4120 &  0.9539 &  3.3222  \\
{\emph{Kurt}}&  4.2276 &    2.5955  &    0.8365  &    2.9065&  3.6626&  4.8803 &   8.8764\\
{\emph{Size}} &   15.6493 &    1.1759  &   13.9694   &  14.8061   &  15.5125 &  16.3595 &  17.7925  \\
{\emph{Bm}} &  1.1330   & 1.0728&   0.2007 &  0.4503 &   0.7892 &   1.4282 &  3.2417   \\
{\emph{Roe}} &  6.9378  &  129.1204 &   -2.5706 &      3.9800 &   8.2984    &  13.1446   &23.2316 \\
{\emph{Lev}}& 47.7206 &  21.6230 &   13.0359   & 32.3652 &     48.7223  & 63.5548  &   78.7176 \\
{\emph{Degree}} &  0.0033 &  0.0045& 0.0010& 0.0010&0.0020& 0.0034&0.0113\\
{\emph{Eigen}}&  0.0312 &  0.0664 & 0.0000  & 0.0009&    0.0077&   0.0309  & 0.1344 \\
{\emph{Between}}  & 0.0026  &  0.0068  & 0.0000   &0.0000 &0.0000 & 0.0022 &   0.0128  \\
 \bottomrule
\end{tabular}
\label{TB:VariableStat}
\end{table}

Table~\ref{TB:VariableCorr} presents the correlation matrix of the variables. The Pearson correlations and the Spearman correlations are presented in the lower and upper triangles. Not surprisingly, there are  positive correlations among \emph{Degree}, \emph{Eigen}, and \emph{Between}, indicating that these three variables play a common role in describing firm positions in the network. Moreover, we find a significant and positive relation between \emph{SYN} and those three network centrality variables. Some of the control variables (e.g., \emph{Size}, \emph{Roe} and \emph{Lev}) are positively correlated with \emph{SYN}, leading us to further investigate the relationship between the position of a firm in the cross-shareholding network and the stock price synchronicity after controlling for these important factors.

\begin{table}[h]
\centering
 \caption{Correlation matrix of the key variables of interest. The Pearson correlations are listed below the diagonal and the Spearman correlations are above the diagonal.}
 \medskip
\footnotesize
 \begin{tabular}{lccccccccccc}
 \toprule
 & $\mathop R\nolimits^2 $ &{\emph{SYN}} &{\emph{Degree}} &{\emph{Eigen}} &{\emph{Between}}   &  {\emph{Skew}}  &   {\emph{Kurt}}   &  {\emph{Size}}   &    {\emph{Bm}}   &   {\emph{Roe}}   &   {\emph{Lev}} \\
\hline
$R^2 $&1.0000 &  1.0000   &0.1588&   0.0877 &  0.0958 & -0.1131 & -0.1244 &  0.2139 &  0.0606 & -0.0663 &  0.0521 \\
{\emph{SYN}} &0.9958   &1.0000   &0.1588  & 0.0877 &  0.0958&  -0.1131 & -0.1244  & 0.2139 &  0.0606&  -0.0663  & 0.0521\\
{\emph{Degree}}&0.1042 &  0.1080  & 1.0000&   0.5811 &  0.8870 &  0.0463 & -0.0918 &  0.1233&   0.1624 &  0.0677 &  0.0835\\
{\emph{Eigen}}& 0.0801 &  0.0820  & 0.7353  & 1.0000&   0.5567&   0.0602  & 0.0171 &  0.1697 &  0.1391&   0.0596  & 0.0945\\
{\emph{Between}}  &0.0897  & 0.0940  & 0.8630   &0.5352  & 1.0000 &  0.0346  &-0.0122  & 0.1994&   0.1607  & 0.0545 &  0.0847 \\
{\emph{Skew}}&-0.0622& -0.0611 &  0.0331  & 0.0331 & -0.0029  & 1.0000  & 0.1357&  -0.0522 &  0.1629&  -0.0354 &  0.0676 \\
{\emph{Kurt}}& -0.1266 & -0.1244 & -0.0265  & 0.0113&  -0.0059&  -0.0409&   1.0000 &  0.0719  & 0.0280 & -0.1351 &  0.0113 \\
{\emph{Size}} & 0.0963  &  0.1078  &  0.1896   & 0.1042   & 0.1621  & -0.0090 &   0.0098  &  1.0000  & -0.1090 &   0.2737 &   0.0932 \\
{\emph{Bm}} &0.0621   &0.0640&   0.1109 &  0.1073  & 0.1033 &  0.1376  & 0.0737 & -0.0004 &  1.0000 & -0.2022  & 0.6273 \\
{\emph{Roe}} &0.0087  & 0.0077 &  0.0126 &  0.0076 &  0.0193 & -0.0257  & 0.0117 &  0.0093&  -0.0211&   1.0000 &  0.0061 \\
{\emph{Lev}}&0.0433  & 0.0445 &  0.0434  & 0.0485 &  0.0437  & 0.0425  & 0.0137  & 0.0053 &  0.5215 & -0.0118 &  1.0000 \\
 \bottomrule
\end{tabular}
\label{TB:VariableCorr}
\end{table}


%

\section{Empirical study}
\label{S1:Empirical}

In this section, univariate analyses and multivariate analyses are employed to examine the relation between cross-shareholding and synchronicity. We also test our results using alternative measures of stock price synchronicity.

\subsection{Univariate analysis}

We begin with the results of \emph{SYN}, which measures the speed at which market-wide and industry-wide information is reflected on the securities market. Because increasing the level of the cross-shareholding reduces information asymmetry between firms and investors, we expect that firms at more central positions in the cross-shareholding network would have fewer irrational pricing factors in the stock market and their stocks would be more properly priced, leading to a higher level of price synchronicity. A competing expectation is that, if a firm is at a more central position in the network, the improvement of the information environment as a result of cross-shareholding would accelerate the generation of firm-specific information, reducing stock price synchronicity \citep{Hutton-Marcus-Tehranian-2009-JFE,Gul-Kim-Qiu-2010-JFE,Bai-Hu-Liu-Zhu-2016-JFS}.

In each fiscal year, we sort the sample during that year into three groups based on the total market value ({\emph{Size}}) at the end of the previous fiscal year. For each size group, we partition the data into five groups according to the sorted contemporaneous return on equity (\emph{Roe}), where each group has the same number of data points. For each {\emph{Size}}-{\emph{Roe}} group, we further partition the data into five groups according to the sorted centrality measure (\emph{Degree}, \emph{Eigen} or \emph{Between}), where each group also has the same number of data points. Finally, $3 \times 5 \times 5$ stock groups based on \emph{Size}, \emph{Roe} and each centrality measure are formed and the equal-weighted average levels of stock price synchronicity are calculated in each \emph{Size-Roe-centrality} group.

\setlength\tabcolsep{3.5pt}
\begin{table}[h]
\caption{$R^2$ for each \emph{Size-Roe-centrality} group.}
\scriptsize
 \smallskip
 \begin{tabular}{ccccccd{1.6}cccccd{2.6}cccccd{2.6}cccccccccccc}
 \toprule
 \multicolumn{19}{l} {Panel A: Small Size}\\
 \hline
&D1&D2&D3&D4&D5&\multicolumn{1}{c}{D5-D1}&E1&E2&E3&E4&E5&\multicolumn{1}{c}{E5-E1}&B1&B2&B3&B4&B5&\multicolumn{1}{c}{B5-B1}\\\hline
Roe1&0.395&0.525&0.575&0.505&0.554&0.1574***&0.509&0.477&0.525&0.507&0.518&0.0089&0.506&0.504&0.480&0.498&0.555&0.0490**$~~$\\
Roe2&0.443&0.562&0.577&0.510&0.502&0.0593**&0.556&0.486&0.525&0.502&0.512&-0.0452*&0.506&0.541&0.520&0.498&0.523&0.0163\\
Roe3&0.465&0.559&0.600&0.513&0.529&0.0644***&0.562&0.529&0.545&0.510&0.521&-0.0416*&0.538&0.527&0.549&0.519&0.530&-0.0079\\
Roe4&0.451&0.508&0.579&0.497&0.522&0.0713***&0.491&0.497&0.526&0.519&0.526&0.0356*&0.516&0.517&0.501&0.496&0.531&0.0151\\
Roe5&0.460&0.534&0.522&0.472&0.512&0.0524**&0.493&0.498&0.518&0.501&0.498&0.0050&0.509&0.495&0.517&0.487&0.501&-0.0076\\
\hline
\multicolumn{19}{l} {Panel B: Median Size}\\
\hline
&D1&D2&D3&D4&D5&\multicolumn{1}{c}{D5-D1}&E1&E2&E3&E4&E5&\multicolumn{1}{c}{E5-E1}&B1&B2&B3&B4&B5&\multicolumn{1}{c}{B5-B1}\\\hline
Roe1&0.495&0.592&0.539&0.586&0.569&0.0531**$~~$&0.560&0.523&0.571&0.570&0.574&0.0148&0.559&0.541&0.578&0.546&0.573&0.0140\\
Roe2&0.474&0.578&0.573&0.548&0.571&0.0444**$~~$&0.524&0.557&0.545&0.551&0.560&0.0363&0.537&0.549&0.518&0.571&0.556&0.0194\\
Roe3&0.458&0.595&0.526&0.593&0.561&0.0664***&0.535&0.540&0.533&0.565&0.555&0.0197&0.541&0.521&0.524&0.582&0.553&0.0117\\
Roe4&0.481&0.565&0.543&0.552&0.558&0.0770***&0.523&0.511&0.566&0.528&0.557&0.0344*&0.508&0.542&0.539&0.538&0.570&0.0622***\\
Roe5&0.465&0.548&0.509&0.527&0.519&0.0557**$~~$&0.515&0.493&0.545&0.495&0.518&0.0025&0.507&0.507&0.522&0.516&0.502&-0.0132\\
\hline
\multicolumn{19}{l} {Panel C: Large Size}\\
\hline
&D1&D2&D3&D4&D5&\multicolumn{1}{c}{D5-D1}&E1&E2&E3&E4&E5&\multicolumn{1}{c}{E5-E1}&B1&B2&B3&B4&B5&\multicolumn{1}{c}{B5-B1}\\\hline
Roe1&0.537&0.618&0.603&0.628&0.639&0.1026***&0.565&0.595&0.605&0.641&0.629&0.0631**$~~$&0.573&0.588&0.599&0.634&0.626&0.0531**$~~$\\
Roe2&0.566&0.597&0.588&0.596&0.619&0.0537**$~~$&0.565&0.601&0.566&0.606&0.628&0.0633**$~~$&0.576&0.597&0.560&0.613&0.621&0.0444**$~~$\\
Roe3&0.522&0.588&0.591&0.603&0.630&0.1085***&0.549&0.571&0.582&0.591&0.630&0.0811***&0.570&0.545&0.585&0.590&0.637&0.0664***\\
Roe4&0.482&0.578&0.554&0.567&0.604&0.1219***&0.532&0.537&0.561&0.551&0.598&0.0666***&0.533&0.538&0.536&0.567&0.610&0.0770***\\
Roe5&0.471&0.501&0.494&0.505&0.569&0.0983***&0.470&0.516&0.518&0.461&0.582&0.1119***&0.485&0.491&0.489&0.520&0.540&0.0557**$~~$\\
\bottomrule
\end{tabular}
\smallskip
\\
This table presents the equal-weighted average of ${R^2}$ for the firm-years in each group and test the hypothesis that ${R^2}$ are equal in the high versus low centrality groups. \emph{D} represents degree centrality, \emph{E} represents Eigenvector centrality, and \emph{B} represents Betweenness centrality. The superscripts *, ** and *** indicate respectively the significance at the 10$\%$, 5$\%$, and 1$\%$ level, respectively.
\label{TB:Groups:R2}
\end{table}

We calculate the equal-weighted average of ${R^2}$ for the firm-years in each group and test the hypothesis that ${R^2}$ is equal in the high versus low centrality groups. The results are presented in Table~\ref{TB:Groups:R2}.
%
For the \emph{Size-Roe-Degree} groups, stock synchronicity quantified by $R^2$ increases with the firm's centrality \emph{Degree} for all \emph{Size-Roe} combinations and the differences between D1 and D5 are statistically significant.
Similar phenomena are also observed  in the \emph{Size-Roe-Eigen} and the \emph{Size-Roe-Between} groups for large-size firms, but not for small-size and medium-size firms. It shows that the impact of cross-shareholding on stock price synchronicity is more likely to be significant in the large-sized groups.
These results imply that the stock prices of large firms tend to mirror the market to a greater extent than those of small firms \citep{Roll-1988-JF,Hutton-Marcus-Tehranian-2009-JFE,Gul-Kim-Qiu-2010-JFE}, and large firms would have fewer insider transactions which improves the incorporation of firm-specific information into individual stock price \citep{Piotroski-Roulstone-2004-AR}.

In addition, the impact of cross-shareholding always exists in different \emph{Size-Roe-Degree} groups. It implies that \emph{Degree} can be more intuitive in describing a firm's position in the cross-shareholding network; however, it is not our focus to discuss the difference between network centrality measures. Lastly, we find that \emph{Roe} would neither strengthen or weaken the relationship between the level of stock price synchronicity and the level of cross-shareholding. We later employ firm characteristic variables as controls in the regression analyses. Overall, our results indicate that the higher level of a firm's cross-shareholding, or the more central its cross-shareholding position in the network, the higher its level of stock price synchronicity.

\subsection{Multivariate analysis}

\interfootnotelinepenalty=10000

In this section, we formalize our analysis using regression to investigate the overall effect of cross-shareholding on stock price synchronicity. We regress stock synchronicity \emph{SYN} against firm centrality $Net$ and control for a range of other variables, where $Net_j$ represents the three centrality measures of firm $j$ in the cross-shareholding network. We also include the log of size at the end of the previous fiscal year (\emph{Size}), the book-to-market ratio at the end of the previous fiscal year (\emph{Bm}), the leverage at the end of the previous fiscal year (\emph{Lev}), and the contemporaneous return on equity (\emph{Roe}) as additional controls for firm characteristics. Larger firms operating in a wider cross-section of the economy are expected to have higher ${R^2}$. Besides, larger firms may be better-covered by media disclosure \citep{Li-Qiao-Zhao-2018-IREF}. The book-to-market ratio places firms along a growth-versus-value spectrum and could thus be systematically related to ${R^2}$. Leverage is also expected to affect ${R^2}$ because it represents the future uncertainty of a firm. Following \cite{Jin-Myers-2006-JFE}, \cite{Hutton-Marcus-Tehranian-2009-JFE}, and \cite{An-Zhang-2013-JCF}, skewness and kurtosis, calculated from Eq.~(\ref{Eq:RegEq})\footnote{We take the residuals from Eq.~(\ref{Eq:RegEq}) and they are highly skewed. We transform them to a roughly symmetric distribution by defining \emph{Residual Return} as the log of one plus the residual return from Eq.~(\ref{Eq:RegEq}). \emph{Skew} and \emph{Kurt} are the skewness and the kurtosis of \emph{Residual Return}, respectively.}, are also considered as control variables in our examination of ${R^2}$ and the level of cross-shareholding in order to assuage the concern that our results are unduly influenced by the distributional properties of residual returns. All variables are standardized to have a mean of zero and a standard deviation of one. Thus, we estimate the following regression equation:
\begin{equation}
{\emph{SYN}}_{j,t} = \alpha_0+\alpha_1 Net_{j,t}+\sum_k {{\gamma_k}Control_j^k}+(YearDummies)+(IndustryDummies)+{\varepsilon_{j,t}},
\label{Eq:MultiVarAnal}
\end{equation}
where we also include the industrial dummies and the year dummies. In the regressions, we use \emph{Degree}, \emph{Eigen} or \emph{Between} for the dependent variable $Net$. We consider three models. Model 1 contains only the explanatory variable $Net$ and the constant term, Model 2 also contains the control variables, and Model 3 contains further the dummy variables.

The regression results are reported in Table~\ref{TB:MultiVarAnal:Models}. Model 1 represents the results without any other control variables except for the firm characteristics including the year fixed effects and the industry fixed effects that are considered to control for macroeconomic changes. The numbers in parentheses represent the $t$-values that are adjusted using standard errors corrected for clustering at the firm level. In Model 1, the coefficients of \emph{Degree}, \emph{Eigen}, and \emph{Between} are 0.1080, 0.0820 and 0.0940, respectively, showing that firms' centrality in the cross-shareholding network is significantly and positively related to stock price synchronicity, even after controlling for other endogenous factors in Model 2 and Model 3. Consistent with Table~\ref{TB:VariableStat}, these results support the idea that there is less information asymmetry arising from a high level of cross-shareholding, and therefore promotes stock price synchronicity \citep{Kelly-2007,Dasgupta-Gan-Gao-2010-JFQA}.
As shown in Model 2 and Model 3, consistent with \cite{Roll-1988-JF}, the coefficients of firm size are significant and positive (0.1320 with $t=6.283$, 0.1350 with $t=6.416$, and 0.1320 with $t=6.321$ for the three centrality measure in Model 3), which supports the notion that a larger firm operates in a wider cross-section of the economy \citep{Hutton-Marcus-Tehranian-2009-JFE}. Furthermore, after controlling the firm size, the results address the problematic feature of a mechanical positive association between firm size and measures of firm connectedness \citep{Larcker-So-Wang-2013-JAE} and a potential concern about the relation (derived from the firm size) between accounting opacity, or media disclosure and stock price informativeness \citep{Easley-Hvidkjaer-Maureen-2002-JF,Li-Qiao-Zhao-2018-IREF}. The book-to-market ratio is significantly and positively correlated with stock price synchronicity. Moreover, both kurtosis and skewness turn out to be statistically significant, where higher values for skewness or kurtosis implying lower levels of stock price synchronicity. This seems reasonable, because extreme events tend to weaken the relationship between firm returns and market-wide information \citep{Hutton-Marcus-Tehranian-2009-JFE}. In addition, the coefficients of the leverage are significant and negative (-0.0966 with $t=-5.058$, -0.0973 with $t=-5.078$, and -0.097 with  $t=-5.069$, respectively, in Model 3) after considering the year and industry dummies, indicating that a higher level of uncertainty for future production and operations deriving from higher levels of the leverage leads to abnormal fluctuations in stock prices. However, the contemporaneous return on equity \emph{Roe} is insignificant in both Model 2 and Model 3. Our results remain unchanged using $Rd$, $Re$ and $Rb$ as the centrality measures, or using firms with fiscal year more than 40 weeks of stock-return data \citep{Chan-Hameed-2006-JFE}. Overall, there is preliminary evidence that a higher centrality in the cross-shareholding network leads to lower information asymmetry, and promotes the reflection of fundamental information in the stock price. We find that the $R$-square is low (0.0120, 0.0070 and 0.0090) in Model 1, which increases about ten times (0.0790, 0.0790 and 0.0780) when the control variables are included in Model 2. More intriguingly, when we consider the year and industry variables, the $R$-square becomes as large as 0.3050.

\setlength\tabcolsep{4.5pt}
\begin{table}[h]
\caption{Relation between stock price synchronicity and centrality in the cross-shareholding network.}
\smallskip
 \footnotesize
 \begin{tabular}{lccccccccc}
  \toprule
 &\multicolumn{3}{c} {Model 1}& \multicolumn{3}{c} {Model 2}& \multicolumn{3}{c} {Model 3}\\
\hline
{\emph{Degree}} & 0.1080***  & &  & 0.0594***   & &  & 0.0286**& &  \\
       &(5.820) & & &(4.003)  && &(2.215)& & \\
{\emph{Eigen}} & &0.0820*** &  & &0.05180*** &  & &0.02760** &  \\
       & &(4.973) & &  &(4.076)& &&(2.327) & \\
{\emph{Between}} & & &0.0940***  & & &0.0440***  && & 0.02840**  \\
       & & &(5.050) &  &&(3.199) && &(2.412) \\
{\emph{Size}} & & & &0.2130***&0.2170***&0.2150***&0.1320***&0.1350***&0.1320***\\
   & & & &(10.980)&(11.040)&(10.990)&(6.283)&(6.416)&(6.321)\\
{\emph{Bm}}   & & & &0.08410***&0.08590***&0.08640***&0.2070***&0.2070***&0.2070***\\
    & & & &(5.090)&(5.176)&(5.205)&(10.360)&(10.350)&(10.360)\\
{\emph{Skew}}& & & &-0.0596***&-0.0592***&-0.05760***&-0.1320***&-0.1320***&-0.1310***\\
       & & & & (-5.304)&(-5.260)&(-5.108)&(-7.031)&(-7.065)&(-7.010)\\
{\emph{Kurt}} & & & &-0.1430***&-0.1450***&-0.1440***&-0.0775***&-0.0779***&-0.0773***\\
& & & &(-10.200)&(-10.430)&(-10.340)&(-5.350)&(-5.390)&(-5.342)\\
{\emph{Roe}} &&&&0.0004&0.0007&0.0003&0.0008&0.0008&0.0006\\
&&&&(0.044)&(0.068)&(0.034)&(0.134)&(0.142)&(0.094)\\
{\emph{Lev}}&&&&-0.0147&-0.0160&-0.0156&-0.0966***&-0.0973***&-0.0970***\\
       &&&&(-0.925)&(-1.004)&(-0.975)&(-5.058)&(-5.078)&(-5.069)\\
Year FE    & & & & & & &Yes&Yes&Yes\\
Industry FE& & & & & & &Yes&Yes&Yes\\
{\emph{Constant}} &0.0000&0.0000&0.0000&0.0000&0.0000&0.0000&4.9090***&4.9130***&4.9110***\\
       &(0.000)&(0.000)&(0.000)&(0.000)&(0.000)&(0.000)&(55.930)&(55.920)&(55.860)\\
Observations &7287 &7287 &7287 &7287 &7287 &7287 &7287 &7287 &7287 \\
$R$-squared &0.0120&0.0070&0.0090&0.0790&0.0790&0.0780&0.3050&0.3050&0.3050\\
\bottomrule
\end{tabular}
\smallskip
\\
This table contains ordinary least square regressions of \emph{SYN} as a function of the centrality in the cross-shareholding network and control variables. \emph{SYN} measures stock price synchronicity: ${\emph{SYN}} = \ln\left[R^2/(1-R^2)\right]$. There are 7287 firm-years in the sample period between 2004-2016. The centrality measures includes \emph{Degree}, \emph{Eigen}, and \emph{Between}. Control variables include the log of size (\emph{Size}), the book-to market ratio (\emph{Bm}) and the leverage (\emph{Lev}) at the end of the previous fiscal year, skewness (\emph{Skew}), kurtosis (\emph{Kurt}), contemporaneous return on equity (\emph{Roe}), and indicator variables for year and the industry classification. All variables are standardized to have zero mean and unit standard deviation. Numbers in parentheses represent $t$-values that are adjusted using standard errors corrected for clustering at the firm level. *, **, *** indicate significance at the 10$\%$, 5$\%$, and 1$\%$ level, respectively.
\label{TB:MultiVarAnal:Models}
\end{table}

Prior research mentioned that the informativeness of insider trades decreases with firm size \citep{Piotroski-Roulstone-2004-AR}. Having observed a positive association between the cross-shareholding and stock price synchronicity, we next consider whether this relation persists across firms with different sizes. Specifically, identifying the size of firms for which this association is particularly strong can provide insights into possible underlying economic mechanisms driving price synchronicity as well as mitigate concerns that the preceding results are simply a consequence of firm size. We stratify our sample into terciles based on annual rankings of end-of-year total market values, and re-estimate Eq.~(\ref{Eq:MultiVarAnal}) by terciles. According to Table~\ref{TB:MultiVarAnal:Models}, we employ Model 3 here.

Table~\ref{TB:MultiVarAnal:SizeSorted} represents the results of the three size groups. The coefficients of \emph{Degree}, \emph{Eigen} and \emph{Between} for the large-size group are 0.0632, 0.0592 and 0.0580, respectively (with a significance of 5$\%$), while the coefficients of the three centrality measures are all insignificant for the small-size and medium-size groups. The impact of cross-shareholding on stock price synchronicity only significantly exists in large firms, indicating that cross-shareholding can reduces information asymmetry more effectively for large firms by constructing long-term and stable links between shareholders, while cross-shareholding does not seem to improve the information environment of small firms (as avenues of short-term finance and arbitrage). These estimations of centrality do not reveal any monotonic trends in the relations across size terciles. We also find that the marginal effect of size on price synchronicity tends to be stronger among small firms, extending the results in previous studies \citep{Roll-1988-JF,Chan-Hameed-2006-JFE,Hutton-Marcus-Tehranian-2009-JFE}. Consistent with the results in Table~\ref{TB:MultiVarAnal:Models}, the book-to-market ratio is significantly and positively correlated with stock price synchronicity across all the size groups. The coefficients of \emph{Lev} within the three size groups are significant at the 10$\%$ level, implying that leverage does have an impact on stock price synchronicity regardless of a firm's size. Our results remain unchanged when we use $Rd$, $Re$ and $Rb$ as the centrality measure.

\setlength\tabcolsep{4.5pt}
\begin{table}[h]
\caption{Effect of cross-shareholding on stock price synchronicity by size terciles}
\smallskip
 \footnotesize
 \begin{tabular}{lccccccccc}
  \toprule
 &\multicolumn{3}{c} {Small size}& \multicolumn{3}{c} {Medium size}& \multicolumn{3}{c} {Large size}\\
\hline
{\emph{Degree}} & -0.0127  & &  & 0.0090   & &  & 0.0630**& &  \\
       &(-0.613) & & &(0.520)   && &(2.565)& & \\
{\emph{Eigen}} & &-0.0122&  & &0.1480 &  & &0.0592*** &  \\
       & &(-0.596) & &  &(0.845)& &&(3.202) & \\
{\emph{Between}} & & &0.0042  & & &-0.0022  && & 0.0580** \\
       & & &(0.252) &  &&(-0.143) && &(2.406) \\
{\emph{Size}}&0.1040***&0.1040***&0.1030***&0.1140***&0.1140***&0.1140***&0.0466&0.0588&0.0498\\
   &(2.817)&(2.833)&(2.788)&(2.638)&(2.635)&(2.637)&(1.132)&(1.437)&(1.200)\\
{\emph{Bm}} &0.1040***&0.1040***&0.1030***&0.2140***&0.2140***&0.2150***&0.2360***&0.2380***&0.2370***\\
   &(3.777)&(3.776)&(3.766)&(6.118)&(6.107)&(6.139)&(7.287)&(7.299)&(7.231)\\
{\emph{Skew}}&-0.1980***&-0.1980***&-0.1990***&-0.1630***&-0.1640***&-0.1630***&-0.0355&-0.0357&-0.0332\\
       &(-6.206)&(-6.180)&(-6.229)&(-4.862)&(-4.865)&(-4.857)&(-1.061)&(-1.069)&(-0.993)\\
{\emph{Kurt}} &-0.1120***&-0.1120***&-0.1120***&-0.1010***&-0.1010***&-0.1010***&-0.0314&-0.0325&-0.0309\\
&(-5.119)&(-5.107)&(-5.107)&(-4.371)&(-4.378)&(-4.389)&(-1.090)&(-1.127)&(-1.075)\\
{\emph{Roe}}&-0.0008&-0.0009&-0.0009&-0.0006&-0.0008&-0.0006&-0.0004&-0.0004&-0.0012\\
&(-0.034)&(-0.037)&(-0.040)&(-0.032)&(-0.042)&(-0.030)&(-0.071)&(-0.063)&(-0.202)\\
{\emph{Lev}}&-0.0895***&-0.0894***&-0.0895***&-0.0888***&-0.0887***&-0.0892***&-0.0573*&-0.0601*&-0.0596*\\
       &(-2.701)&(-2.695)&(-2.697)&(-3.545)&(-3.539)&(-3.561)&(-1.802)&(-1.884)&(-1.872)\\
Year FE &Yes&Yes&Yes&Yes&Yes&Yes&Yes&Yes&Yes\\
Industry FE&Yes&Yes&Yes&Yes&Yes&Yes&Yes&Yes&Yes\\
{\emph{Constant}}&5.2250***&5.2260***&5.2300***&1.0640**&1.0670**&1.0640**&0.4410***&0.4230***&0.4580***\\
      &(37.270)&(37.340)&37.360)&(2.144)&(2.147)&(2.136)&(4.045)&(3.931)&(4.308)\\
Observations &2425 &2425 &2425 &2432 &2432 &2432 &2430 &2430 &2430 \\
$R$-squared &0.3270&0.3270&0.3270&0.3230&0.3230&0.3230&0.3150&0.3150&0.3150\\
\bottomrule
\end{tabular}
\smallskip
\\
This table shows the effect of cross-shareholding on stock price synchronicity by size terciles. \emph{SYN} measures stock price synchronicity: ${\emph{SYN}} = \ln\left[R^2/(1-R^2) \right]$. There are 7287 firm-years in the sample period between 2004-2016. The centrality measures includes \emph{Degree}, \emph{Eigen}, and \emph{Between}. Control variables include the log of size (\emph{Size}), the book-to market ratio (\emph{Bm}) and the leverage (\emph{Lev}) at the end of the previous fiscal year, skewness (\emph{Skew}), kurtosis (\emph{Kurt}), contemporaneous return on equity (\emph{Roe}), and indicator variables for year and the industry classification. All variables are standardized to have zero mean and unit standard deviation. Numbers in parentheses represent $t$-values that are adjusted using standard errors corrected for clustering at the firm level. *, **, *** indicate significance at the 10$\%$, 5$\%$, and 1$\%$ level, respectively.
\label{TB:MultiVarAnal:SizeSorted}
\end{table}

\subsection{Robustness}

In this section, we test if our inferences remain unchanged by using alternative measures of price synchronicity. Moreover, to rule out the possibility that our regression results are driven by outliers, we re-run the analysis using a winsorized sample.

In emerging markets, including industry returns as an additional factor to explain stock returns in a market model is problematic because in some markets the economy is dominated by a few industries and it is therefore difficult to disentangle the industry effects from the market effects \citep{Chan-Hameed-2006-JFE}, which is also the case for the Chinese stock market \citep{Han-Xie-Xiong-Zhang-Zhou-2017-FNL}. In addition, it is common for an industry in an emerging economy to include only a few companies. Consequently, when the industry returns are computed using the few companies from the given industry, they reflect firm-specific news rather than industry-wide news.
To reconcile this potential concern, we employ the market model \citep{Chan-Hameed-2006-JFE,Nguyen-Truong-2013-JIFMIM}\begin{equation}
 {r_{j,t}} = {\alpha _j} + {\beta _j}{r_{m,t}} + {{\rm{\varepsilon }}_{j,t}}
 \label{Eq:MarketModel}
\end{equation}
as well as the market-industry model without non-synchronous trading \citep{Crawford-Jones-Roulstone-So-2012-AR,Li-Qiao-Zhao-2018-IREF}
\begin{equation}
 {r_{j,t}} = {\alpha _j} + {\beta _{1,j}}{r_{m,t}} + {\beta _{2,j}}{r_{i,t}} + {\varepsilon _{j,t}}
 \label{Eq:MarketIndustryModel}
\end{equation}
to calculate the stock price synchronicity, noting them as \emph{SYN}1 and \emph{SYN}2, respectively. Therefore, we run these two regressions in Model 4 and Model 5. In Model 6, we winsorize all variables at the bottom and top 1$\%$ of their empirical distributions.
In Model 7, we also exclude the years 2008, 2015 and 2016 from our tests to avoid potential bias that can arise from the unusual price movements during the years of the financial crisis and the stock market crash, and our conclusions remain unaffected\footnote{We use standard errors corrected for clustering at the year level with bootstrapped (1000 replications) standard errors to drive out a potential concern that standard errors are downwards biased with small numbers of clusters \citep{Drake-Jennings-Roulstone-2017-MS}. Furthermore, as \cite{Kelly-2007} posits that the small and the young firms keep from informed trading, which brings down stock price synchronicity, we also include firms' age (\emph{Age}), the date since firms' first appearance on the Shanghai or the Shenzhen exchanges, as control variables, and from the unreported results we can resolve a potential selection bias caused by size and age.}.

\setlength\tabcolsep{3pt}
\begin{table}[h]
\caption{Robustness test.}
\smallskip
\scriptsize
 \begin{tabular}{lcccccccccccc}
 \toprule
 &\multicolumn{3}{c} {Model 4}& \multicolumn{3}{c} {Model 5}& \multicolumn{3}{c} {Model 6}& \multicolumn{3}{c} {Model 7}\\
\hline
{\emph{Degree}} &0.0227**  & &  & 0.0322***   & &  & 0.0281*& &  &0.0264*  &  &  \\
       &(2.096)    & & & (2.668)      && &(1.936)  & & &(1.823) & & \\
{\emph{Eigen}} & &0.0244***&  & &0.0325***&  & &0.0275** &  &  &0.0272**  &  \\
       & &(2.649) & &  &(2.889)& &&(2.212) &  & &(2.146)& \\
{\emph{Between}} & & &0.0136  & & &0.0324***  && & 0.0303**&&& 0.0230*\\
       & & &(1.209) &  &&(2.990) && &(2.229) &&&(1.748)\\
{\emph{Size}}&0.0476***&0.0501***&0.0506***&0.1420***&0.1460***&0.1430***&0.1570***&0.1600***&0.1560***&0.1470***&0.1500***&0.1480***\\
   &(2.703) &(2.882) &(2.892) &(7.139) &(7.288) &(7.171) &(7.851) &(8.045) &(7.850)&(6.267)&(6.369)&(6.312)\\
{\emph{Bm}} &0.1880***&0.1880***&0.1890***&0.2070***&0.2070***&0.2070***&0.1980***&0.1980***&0.1980***&0.2260***&0.2270***&0.2270***\\
   &(10.52)&(10.53)&(10.570)&(10.040)&(10.030)&(10.040)&(9.946)&(9.949)&(9.923)&(10.410)&(10.430)&(10.430)\\
{\emph{Skew}}&0.0113&0.0114&0.0115&-0.0451***&-0.0453***&-0.0451***&-0.1400***&-0.1410***&-0.1400***&-0.1660***&-0.1670***&-0.1660***\\
       &(0.762) &(0.765) &(0.772) &(-3.662) &(-3.679) &(-3.657) &(-6.825) &(-6.856) &(-6.810)&(-7.756)&(-7.786)&(-7.738)\\
{\emph{Kurt}} &-0.1170***&-0.1170***&-0.1170***&-0.1470***&-0.1470***&-0.1460***&-0.0814***&-0.0820***&-0.0811***&-0.0701***&-0.0704***&-0.0700***\\
&(-8.287)&(-8.323)&(-8.277)&(-11.450)&(-11.570)&(-11.390)&(-5.808)&(-5.875)&(-5.783)&(-4.405)&(-4.424)&(-4.400)\\
{\emph{Roe}}&0.0011&0.0011&0.0010&0.0001&0.0001&-0.0002&-0.0891***&-0.0899***&-0.0892***&0.0016&0.0017&0.0014\\
&(-0.034)&(-0.037)&(-0.040)&(-0.032)&(-0.042)&(-0.030)&(-0.071)&(-0.063)&(-0.202)&(0.258)&(0.267)&(0.225)\\
{\emph{Lev}}&-0.0221&-0.0225&-0.0229&-0.0972***&-0.0980***&-0.0976***&-0.1040***&-0.1050***&-0.1040***&-0.1050***&-0.1050***&-0.1050***\\
       &(-1.517)&(-1.549)&(-1.572)&(-4.669)&(-4.693)&(-4.680)&(-6.308)&(-6.345)&(-6.299)&(-4.810)&(-4.822)&(-4.821)\\
{\emph{Constant}}&-1.5190***&-1.5240***&-1.5220***&4.2660***&4.2690***&4.2650***&3.7700***&3.7730***&3.7740***&4.7070***&4.7100***&4.7080***\\
      &(-16.540) &(-16.570) &(-16.560) &(57.830) &(58.090) &(57.610) &(38.760) &(38.770) &(38.730)&(60.870)&(61.130)&(60.610)\\
Year FE &Yes&Yes&Yes&Yes&Yes&Yes&Yes&Yes&Yes&Yes&Yes&Yes\\
Industry FE&Yes&Yes&Yes&Yes&Yes&Yes&Yes&Yes&Yes&Yes&Yes&Yes\\
Observations &7287 &7287 &7287&7287&7287&7287&7287&7287&7287&6202&6202&6202\\
$R$-squared &0.3450&0.3450&0.3450&0.3290&0.3290&0.3290&0.3120&0.3120&0.313&0.2310&0.2310&0.2310\\
\bottomrule
\end{tabular}
\smallskip
\\
This table shows the robustness tests. In Model 4 and Model 5, the dependent variable are \emph{SYN}1 and \emph{SYN}2, calculated from Eq.~(\ref{Eq:MarketModel}) and Eq.~(\ref{Eq:MarketIndustryModel}), respectively. Variables are all winsorized at the bottom and top 1$\%$ of their empirical distributions in Model 6. Years when the market experiences extreme price movement are excluded in Model 7. All variables are standardized to have a mean of zero and a standard deviation of one. The centrality measures include \emph{Degree}, \emph{Eigen}, and \emph{Between}. Control variables include the log of size at the end of the previous fiscal year (\emph{Size}) and book-to market ratio at the end of the previous fiscal year (\emph{Bm}), skewness (\emph{Skew}), kurtosis (\emph{Kurt}), contemporaneous return on equity (\emph{Roe}), leverage at the end of the previous fiscal year (\emph{Lev}), and indicator variables for year and the industry classification. Numbers in parentheses represent t-values that are adjusted using standard errors corrected for clustering at the firm level. *, **, *** indicate significance at the 10$\%$, 5$\%$, and 1$\%$ level, respectively.
\label{TB:MultiVarAnal:Robust}
\end{table}

In Table~\ref{TB:MultiVarAnal:Robust}, most of the coefficients of \emph{Degree}, \emph{Eigen}, and \emph{Between} are statistically significant and positive in Model 4 (0.0227, $t=2.096$; 0.0244, $t=2.649$; 0.0136, $t=1.209$, respectively) and in Model 5 (0.0322, $t=2.668$; 0.0325, $t=2.889$; 0.0324, $t=2.990$, respectively). Consistent with our conjecture, the empirical results suggest that centrality in the cross-shareholding network is positively related to stock price synchronicity. In addition, we find that our main regression results are not driven by outliers in Model 6 nor by the turbulent years in Model 7. However, removing the turbulent years reduces the $R$-squared remarkably, though the results are qualitatively the same. This finding reflects the fact that stock prices are more synchronous during market bubbles and especially crashes \citep{Plerou-Gopikrishnan-Gabaix-Stanley-2002-PRE,Billio-Getmansky-Lo-Pelizzon-2012-JFE,Ren-Zhou-2014-PLoS1}.

Cross-sectional dependence would result in a biased estimation of our regressions. An unreported table describes the results of the effect of cross-shareholding on stock price synchronicity with the \cite{Fama-MabBeth-1973-JPE} method including industry fixed effects, based on the heteroskedasticity and autocorrelation consistent standard errors of \cite{Newey-West-1987-Em}. We capture the serial correlation in the estimated coefficients using a first-order autoregressive process. Not surprisingly, the unreported results remain similar.

\section{Further discussion}
\label{S1:Discuss}

\subsection{Effects of cross-shareholding on noise trading}

In the empirical analysis above, a basic conclusion is that the level of cross-shareholding promotes stock price synchronicity, which can be interpreted as firms with higher levels of cross-shareholding, or that are more central in the cross-shareholding network would have higher-quality of information disclosure and information environment, which further improves price informativeness. Nevertheless, firms at more central positions in the cross-shareholding network would reflect more market-wide information, which increases its level of systematic risk. Therefore, it casts reasonable doubt on whether it is through the improvement of the information environment that cross-shareholding promote stock price synchronicity.

Evidence has been given about the impact of the information environment on noise trading. For example, \cite{Roll-1988-JF} suggests that the volatility of stock returns would not be totally dependent on fundamental information. The noise trading has been proven to be a potential factor of return volatility \citep{Anbo-Pantzalis-Park-2017-JBF}. It is shown that taking noise trading into consideration, stock price synchronicity does not always vary monotonically with price informativeness \citep{Lee-Liu-2011-JBF}. In terms of the corporate structure, it is confirmed that managers tend to disclose the good news and hide the bad news to protect their own professional prestige, and such low quality of information disclosure may cause heterogeneous beliefs among investors, which consequently increases the volatility of individual stock return \citep{Jiang-Xu-Yao-2009-JFQA}. Therefore, in this section, we intend to prove the impact of cross-shareholding on stock pricing efficiency by testing the relation between the improvement of the information environment (arising from cross-shareholding) and noise trading.

The dependent variable we are interested in is \emph{DEV}. Specifically, following \cite{Brockman-Yan-2009-JBF}, we define \emph{DEV} as the standard deviation of the log of one plus the residual return from Eq.~(\ref{Eq:RegEq}). The centrality measures include \emph{Degree}, \emph{Eigen} and \emph{Between}. Following \cite{Gu-Kang-Xu-2018-JBF}, we include the log of size at the end of the previous fiscal year (\emph{Size}) and book-to market ratio  at the end of the previous fiscal year (\emph{Bm}). We also include contemporaneous return on equity (\emph{Roe}), leverage at the end of the previous fiscal year (\emph{Lev}), and indicator variables for the year and the industry classification as control variables. Hence, the regression equation reads
\begin{equation}
{\emph{DEV}}_{j,t} = \alpha_0+\alpha_1 Net_{j,t}+\sum_k {{\gamma_k}Control_j^k}+(YearDummies)+(IndustryDummies)+{\varepsilon_{j,t}}.
\label{Eq:MultiVarAnal:NoiseTrading}
\end{equation}
Due to the speculative feature of stock trading in China \citep{Truong-2011-JIFMIM}, as discussed above, we expect firms with a more central position in the network to have lower idiosyncratic volatility after explained by market and industry information.

\setlength\tabcolsep{3pt}
\begin{table}[h]
\caption{Centrality in the cross-shareholding network and the noise trading.}
\smallskip
\scriptsize
 \begin{tabular}{lcccccccccccc}
 \toprule
 &\multicolumn{3}{c} {All sample}& \multicolumn{3}{c} {Small size}&\multicolumn{3}{c} {Medium size}&\multicolumn{3}{c} {Large size}\\
\hline
{\emph{Degree}}&-0.0282**& & &-0.0149& & &-0.0245& & &-0.0589***& & \\
      &(-2.579)& & &(-0.709)& & &(-1.410)& & &(-3.344)& & \\
{\emph{Eigen}}&&-0.0207** & &&0.0040 & &&-0.0282* & &&-0.0378** &\\
   &&(-2.106) & &&(0.197) & &&(-1.650) & &&(-2.305)&\\
{\emph{Between}}&& & -0.0260***&& &-0.0258 && &-0.0116 && &-0.0521***\\
&& & (-2.743)&& &(-1.548) && &(-0.633) && &(-3.436)\\
{\emph{Size}}&-0.2140***&-0.2180***&-0.2150***&-0.1240***&-0.1250***&-0.1230***&-0.1310***&-0.1310***&-0.1310***&-0.1400***&-0.1540***&-0.1430***\\
&(-14.730)&(-15.220)&(-14.810)&(-2.886)&(-2.905)&(-2.874)&(-3.294)&(-3.286)&(-3.286)&(-5.335)&(-5.845)&(-5.408)\\
{\emph{Bm}}&-0.1420***&-0.1430***&-0.1430***&-0.1330***&-0.1340***&-0.1330***&-0.1700***&-0.1700***&-0.1710***&-0.1440***&-0.1470***&-0.1450***\\
&(-8.578)&(-8.618)&(-8.585)&(-5.493)&(-5.492)&(-5.519)&(-5.692)&(-5.687)&(-5.710)&(-5.922)&(-6.0120)&(-5.929)\\
{\emph{Roe}}&0.0000&0.0000&0.0002&-0.0018&-0.0019&-0.0018&-0.0173&-0.0169&-0.0175&0.0049&0.0050&0.0057\\
&(-0.001)&(-0.006)&(0.031)&(-0.057)&(-0.062)&(-0.057)&(-1.032)&(-1.019)&(-1.046)&(0.783)&(0.771)&(0.988)\\
{\emph{Lev}}&0.0681***&0.0691***&0.0686***&0.0338&0.0339&0.0341&0.0726***&0.0727***&0.0731***&0.1240***&0.1280***&0.1270***\\
&(3.775)&(3.811)&(3.793)&(1.359)&(1.362)&(1.368)&(2.870)&(2.875)&(2.885)&(5.563)&(5.749)&(5.655)\\
{\emph{Constant}}&-2.1900***&-2.1970***&-2.1910***&-1.3740***&-1.3670***&-1.3710***&-0.5190***&-0.5250***&-0.5170***&-1.0000***&-1.0180***&-1.0180***\\
&(-38.340)&(-38.750)&(-38.260)&(-11.860)&(-11.760)&(-11.830)&(-3.993)&(-4.050)&(-4.004)&(-11.640)&(-11.440)&(-12.240)\\
Year FE&Yes&Yes&Yes&Yes&Yes&Yes&Yes&Yes&Yes&Yes&Yes&Yes\\
Industry FE &Yes&Yes&Yes&Yes&Yes&Yes&Yes&Yes&Yes&Yes&Yes&Yes\\
Observations &7287&7287&7287&7287&7287&7287&7287&7287&7287&7287&7287&7287\\
$R$-squared&0.4110&0.4110&0.4110&0.3250&0.3240&0.3250&0.4180&0.4180&0.4170&0.5200&0.5190&0.5200\\
\bottomrule
\end{tabular}
\smallskip
\\
The dependent variable \emph{DEV} is the standard deviation of the log of one plus the residual return calculated from Eq. (\ref{Eq:RegEq}). The centrality measures includes \emph{Degree}, \emph{Eigen}, and \emph{Between}. Control variables include the log of size at the end of the previous fiscal year (\emph{Size}) and book-to market ratio at the end of the previous fiscal year (\emph{Bm}), contemporaneous return on equity (\emph{Roe}), leverage at the end of the previous fiscal year (\emph{Lev}), and indicator variables for year and the industry classification. All variables are standardized to have a mean of zero and a standard deviation of one. Numbers in parentheses represent $t$-values that are adjusted using standard errors corrected for clustering at the firm level. *, **, *** indicate significance at the 10$\%$, 5$\%$, and 1$\%$ level, respectively.
\label{TB:MultiVarAnal:NoiseTrading}
\end{table}

Table~\ref{TB:MultiVarAnal:NoiseTrading} describes the results. The numbers in parentheses represent the $t$-values that are adjusted using standard errors corrected for clustering at the firm level. The coefficients of the centrality measures are negative and significant at the 5$\%$ level: the coefficients of {\emph{Degree}}, {\emph{Eigen}} and {\emph{Between}} are -0.0282, -0.0207 and -0.0454, respectively, with $t$-statistics of -2.579, -2.106 and -3.926, respectively in the entire sample. The coefficients of the three centrality measures are also negative and significant after controlling for the year fixed effects and the industry fixed effects. The results indicate that the improvement of the information environment, arising from cross-shareholding, would reduce noise trading and thus promotes stock price informativeness \citep{Lee-Liu-2011-JBF}. Therefore, the stock price would reflect more informed trading instead of irrational factors.

Not surprisingly, the coefficient of size is positive and highly significant, implying that given a greater amount of available fundamental information (as compared to small firms), large firms have lower volatility during the year \citep{Gul-Kim-Qiu-2010-JFE,An-Zhang-2013-JCF}. The book-to-market ratio is statistically significant in reducing idiosyncratic volatility. Leverage is associated with higher idiosyncratic volatility since high levels of uncertainty with respect to future cash flow dominates the stock return. Our results remain unchanged using \emph{Rd}, \emph{Re}, and \emph{Rb} as centrality measures.

To investigate how the impact of cross-shareholding on noise trading varies across firms with different sizes, we stratify our sample into terciles based on annual rankings of end-of-year total market values, and re-estimate Eq.~(\ref{Eq:MultiVarAnal:NoiseTrading}) by tercile. The coefficients of \emph{Degree}, \emph{Eigen}, and \emph{Between} are statistically significant and negative in the large-sized groups (-0.0589, $t=-3.344$; -0.0378, $t=-2.305$; and -0.0521, $t=-3.436$, respectively). Consistent with our results in Section~\ref{S1:Empirical}, the results confirm that a higher level of cross-shareholding in large firms would foster more stable relations between shareholders and consequently demands a higher quality of the information environment, while a higher level of cross-shareholding in small firms would be more a reflection of the short-term arbitrage under particular market conditions or demands for temporary financing. We also notice that the $R$-squared value increases significantly from the small-size group to the large-size group.

Another measure of idiosyncratic volatility, mentioned in \cite{Nartea-Wu-Liu-2013-JIFMIM} and \cite{Gu-Kang-Xu-2018-JBF}, is also employed to test the relation between cross-shareholding and noise trading. We calculate the idiosyncratic volatility (\emph{DEV-FF3}) of yearly stock returns as the standard deviation of weekly excess returns relative to the \cite{Fama-French-1993-JFE} three-factor during the same year. More specifically, stock \emph{j}'s idiosyncratic volatility (\emph{DEV-FF3}) in the same year is calculated as below:
\begin{equation}
 r_{j,t} - r_{f,t} = \alpha_j +\beta_j{\emph{MKT}}_t + {s_j}{\emph{SMB}}_t + {h_j}{\emph{HML}}_t + \varepsilon_{j,t},
\end{equation}
where ${r_{j,t}} - {r_{f,t}}$ is stock $j$'s weekly excess return at week \emph{t} and ${\emph{MKT}}_t$, ${\emph{SMB}}_t$ and ${\emph{HML}}_t$ are weekly returns of the \cite{Fama-French-1993-JFE} three factors extracted from the RESSET Financial Database\footnote{Of the 7287 firm-years, 17 were eliminated because of unavailable stock-return data.}.

Table~\ref{TB:MultiVarAnal:NoiseTrading2} shows the regression results. The coefficients of \emph{Degree}, \emph{Eigen} and \emph{Between} are -0.0245, -0.0172, and -0.0210, respectively, with $t$-statistics of -2.240, -1.754 and -2.063, respectively across the whole sample. \emph{Size} is significantly and negatively related to \emph{DEV-FF3}, with coefficients of -0.1200, -0.1240, and -0.1220 in Column 1, Column 2, and Column 3, respectively. Consistent with our results in Table~\ref{TB:MultiVarAnal:NoiseTrading}, cross-shareholding would promote stock pricing efficiency by driving noise trading out, and irrational trading would be more clustered at the small firms. When we turn to investigate how such effects vary across firms of different sizes, the results also show a significant and positive relation between cross-shareholding and noise trading in large firms. The results may be rested on an idea that, compared to small firms (whose goals primarily relate to capital operation and short-term financing), large firms have stable cash flows and lower sensitivity to market-wide shocks, allowing them to be partly owned over the long term by shareholders pursuing profit through continuous cash flows instead of stock price fluctuation, which makes those shareholders demand a better quality of the information environment\footnote{Following \cite{Gu-Kang-Xu-2018-JBF}, we include the yearly turnover ratio of the tradable shares for the present year (\emph{Turnover}), obtained from the RESSET Financial Database, as control variables to measure the liquidity in unreported tests and our unreported results remain unaffected.}.

\setlength\tabcolsep{3pt}
\begin{table}[h]
\caption{Alternative measures of the idiosyncratic volatility.}
\smallskip
\scriptsize
 \begin{tabular}{lcccccccccccc}
 \toprule
 &\multicolumn{3}{c} {All sample}& \multicolumn{3}{c} {Small size}&\multicolumn{3}{c} {Medium size}&\multicolumn{3}{c} {Large size}\\
\hline
{\emph{Degree}}&-0.0245**& & &-0.0170& & &-0.0325**& & &-0.0580***& & \\
      &(-2.240)& & &(-0.784)& & &(-1.981)& & &(-3.215)& & \\
{\emph{Eigen}}&&-0.0172* & &&0.0048 & &&-0.0362** & &&-0.0300* &\\
   &&(-1.754) & &&(0.235) & &&(-2.189) & &&(-1.705)&\\
{\emph{Between}}&& & -0.0210**&& &-0.0248 && &-0.0189 && &-0.0513***\\
&& & (-2.063)&& &(-1.388) && &(-1.126) && &(-2.952)\\
{\emph{Size}}&-0.1200***&-0.1240***&-0.1220***&-0.1500***&-0.1510***&-0.1500***&-0.0901**&-0.0897**&-0.0897**&-0.0301&-0.0458**&-0.0337\\
&(-8.505)&(-9.023)&(-8.673)&(-2.848)&(-2.869)&(-2.843)&(-2.309)&(-2.299)&(-2.298)&(-1.432)&(-2.255)&(-1.618)\\
{\emph{Bm}}&-0.1200***&-0.1210***&-0.1210***&-0.1160***&-0.1170***&-0.1160***&-0.1380***&-0.1380***&-0.1390***&-0.1220***&-0.1260***&-0.1240***\\
&(-7.993)&(-8.044)&(-8.039)&(-4.885)&(-4.895)&(-4.908)&(-5.420)&(-5.410)&(-5.456)&(-5.611)&(-5.801)&(-5.753)\\
{\emph{Roe}}&0.0039&0.0039&0.0041&0.0079&0.0077&0.0079&-0.0060&-0.0056&-0.0063&0.00432&0.0044&0.0051\\
&(0.566)&(0.559)&(0.600)&(0.287)&(0.280)&(0.286)&(-0.363)&(-0.339)&(-0.384)&(0.487)&(0.489)&(0.609)\\
{\emph{Lev}}&0.0589***&0.0599***&0.0595***&0.0305&0.0306&0.0308&0.0543**&0.0545**&0.0548**&0.13300***&0.137***&0.1350***\\
&(3.518)&(3.554)&(3.540)&(1.262)&(1.266)&(1.273)&(2.300)&(2.305)&(2.318)&(6.028)&(6.221)&(6.136)\\
{\emph{Constant}}&-0.4130*&-0.4220*&-0.4080*&1.2620***&1.270***&1.266***&0.688***&0.680***&0.682***&0.578***&0.541***&0.574***\\
&(-1.921)&(-1.951)&(-1.912)&(8.770)&(8.801)&(8.782)&(3.150)&(3.153)&(3.134)&(7.040)&(6.655)&(7.039)\\
Year FE&Yes&Yes&Yes&Yes&Yes&Yes&Yes&Yes&Yes&Yes&Yes&Yes\\
Industry FE &Yes&Yes&Yes&Yes&Yes&Yes&Yes&Yes&Yes&Yes&Yes&Yes\\
Observations &7270&7270&7270&2419&2419&2419&2430&2430&2430&2421&2421&2421\\
R-squared&0.3680&0.3680&0.3680&0.3070&0.306&0.3070&0.3830&0.3830&0.3820&0.4820&0.4800&0.4810\\
\bottomrule
\end{tabular}
\smallskip
\\
Note: The dependent variable, \emph{DEV-FF3}, is the standard deviation of weekly excess returns relative to the \cite{Fama-French-1993-JFE} three-factor during the same year. The centrality measures includes \emph{Degree}, \emph{Eigen} and \emph{Between}. Control variables include the log of size at the end of the previous fiscal year (\emph{Size}) and book-to market ratio at the end of the previous fiscal year (\emph{Bm}), contemporaneous return on equity (\emph{Roe}), leverage at the end of the previous fiscal year (\emph{Lev}), and indicator variables for year and the industry classification. All variables are standardized to have zero mean and unit standard deviation. Numbers in parentheses represent $t$-values that are adjusted using standard errors corrected for clustering at the firm level. *, **, *** indicate significance at the 10$\%$, 5$\%$, and 1$\%$ level, respectively.
\label{TB:MultiVarAnal:NoiseTrading2}
\end{table}

To summarize, consistent with \cite{Kelly-2007} and \cite{Lee-Liu-2011-JBF}, the obvious conclusion can be drawn from the results in Table~\ref{TB:MultiVarAnal:NoiseTrading} and Table~\ref{TB:MultiVarAnal:NoiseTrading2} that the improvement in the quality of the information environment, arising from cross-shareholding, reduces noise trading and further promotes stock price informativeness, especially for large firms.  We again observe that the $R$-squared value increases significantly from the small-size group to the large-size group.

\subsection{Price delay}

In the previous analysis, we provide evidence that cross-shareholding promotes stock price synchronicity through a noise-reducing process. One can expect that if a stock is priced more rationally, the stock returns would reflect more contemporaneous market-wide information, and vice versa. Thus, one testable inference is that the explanatory power of historical information in the stock market would be reduced for firms at more central positions in the cross-shareholding network.

Following \cite{Hou-Moskowitz-2005-RFS}, we measure market-wide information using historical market returns or the average historical returns of stocks in the same industry. If noise trading causes price delay, we should expect this price delay or lead-lag effects in returns to increase as the centrality in the cross-shareholding network decreases, after controlling for other firm characteristics. In addition, \cite{Dong-Li-Lin-Ni-2016-JFQA} reckon that stock price synchronicity, as a measure of price informativeness, is not without criticism.

To measure price delay, we employ measures used in \cite{Hou-Moskowitz-2005-RFS}. The market return or the industry return represents the news to which stocks respond. At the end of December for each year, we run a regression of each stock's weekly return contemporaneous and four weeks of lagged returns of either the value-weighted market portfolio or the value-weighted industry portfolio:
\begin{equation}
 {r_{j,t}} = {\alpha_j} + {\beta_j}{r_{p,t}} + \sum_{k = 1}^4 {\delta^{\left(- k\right)}}{r_{p,t - k}} + {\varepsilon _{j,t}},
 \label{Eq:PriceDelay}
\end{equation}
where ${r_{j,t}}$ denotes the weekly return of firm $j$ at week $t$ and ${r_{p,t}}$ is the return on the value-weighted market index or the value-weighted return of the stocks in the same industry portfolio as stock $j$ at week $t$.  The industry portfolio of a stock is defined as all other stocks in the CSRC with the same industry classification. If the stock contains a larger rational pricing component and responds immediately to market news, $\beta_j$ will be significantly different from zero, but none of ${\delta ^{( - k)}}$ will differ from zero. If stock $j$'s price is followed by much noise trading, some $\delta ^{(- k)}$ will differ significantly from zero. The measure of delay that we use is the fraction of the variation in contemporaneous individual stock returns explained by lagged market or industry returns. The delay measure is thus defined as one minus the ratio of the ${R^2}$ from Eq.~(\ref{Eq:PriceDelay}) restricting ${\delta ^{(- k)}} = 0$, for all $k \in \{ 1, 2, 3, 4\}$, over the ${R^2}$ from Eq.~(\ref{Eq:PriceDelay}) without any restrictions:
\begin{equation}
 {\emph{Delay}} = 1 - \frac{{R_{{\delta ^{\left( { - k} \right)}} = 0,\forall {\rm{k}} \in \left\{ {1,2,3,4} \right\}}^2}}{{{R^2}}}.
 \label{Eq:Delay}
\end{equation}
We start with baseline regressions to investigate the overall effect of cross-shareholding on price delay. To examine the size effect, we stratify our sample into terciles based on the annual rankings of the end-of-year total market values and within each size tercile we regress the price delay measure on the three centrality measures of the cross-shareholding network and other stock characteristics. The measure of delay is bounded between zero and one. Therefore, for the regressions to be well-specified, we include the logistic transformation of the measure of delay as a dependent variable. Following \cite{Pareek-2012}, we include the log of size at the end of the previous fiscal year (\emph{Size}), book-to-market ratio at the end of the previous fiscal year (\emph{Bm}), contemporaneous return on equity (\emph{Roe}), leverage at the end of the previous fiscal year (\emph{Lev}), and indicator variables for the year and industry classification.

\begin{table}[h]
\caption{Centrality in the cross-shareholding network and price delay.}
\smallskip
\scriptsize
 \begin{tabular}{lcccccccccccc}
  \toprule
 &\multicolumn{3}{c} {All sample}& \multicolumn{3}{c} {Small size}&\multicolumn{3}{c} {Medium size}&\multicolumn{3}{c} {Large size}\\
\hline
{\emph{Degree}}&-0.0162& & &0.0164& & &-0.0072& & &-0.0678***& & \\
      &(-1.405)& & &(1.034)& & &(-0.421)& & &(-2.941)& & \\
{\emph{Eigen}}&&-0.0185 & &&0.0144 & &&-0.0045& &&-0.0653*** &\\
   &&(-1.381) & &&(0.865) & &&(-0.237)& &&(-2.612)&\\
{\emph{Between}}&& & -0.0087&& &0.0035&& &0.0007 && &-0.0451**\\
&& & (-0.887)&& &(0.239) && &(0.046) && &(-2.380)\\
{\emph{Size}}&-0.0656***&-0.0670***&-0.0679***&-0.0693**&-0.0697**&-0.0686**&-0.1120***&-0.1120***&-0.1120***&0.0371&0.0244&0.0279\\
&(-4.293)&(-4.484)&(-4.458)&(-2.035)&(-2.047)&(-2.014)&(-2.733)&(-2.731)&(-2.732)&(1.315)&(0.912)&(0.991)\\
{\emph{Bm}}&-0.1170***&-0.1170***&-0.1180***&-0.0976***&-0.0979***&-0.0970***&-0.1400***&-0.1400***&-0.1410***&-0.1080***&-0.1090***&-0.1110***\\
&(-7.215)&(-7.200)&(-7.273)&(-4.052)&(-4.052)&(-4.035)&(-5.382)&(-5.387)&(-5.399)&(-3.144)&(-3.193)&(-3.261)\\
{\emph{Roe}}&-0.0090*&-0.0091*&-0.0090*&0.0068&0.0069&0.0069&-0.0141&-0.0140&-0.0141&-0.0184**&-0.0185**&-0.0177**\\
&(-1.685)&(-1.681)&(-1.684)&(0.538)&(0.547)&(0.551)&(-0.823)&(-0.822)&(-0.826)&(-2.096)&(-2.049)&(-2.088)\\
{\emph{Lev}}&0.0350**&0.0352**&0.0356**&0.0368&0.0366&0.0366&0.0461*&0.0463**&0.0464**&0.0251&0.0279&0.0289\\
&(2.283)&(2.302)&(2.316)&(1.572)&(1.564)&(1.566)&(1.956)&(1.964)&(1.968)&(0.669)&(0.748)&(0.773)\\
{\emph{Constant}}&0.3200***&0.3190***&0.3160***&0.3830***&0.3800***&0.3770***&-0.1670&-0.1680&-0.1670&0.2940***&0.3160***&0.2520**\\
&(4.205)&(4.207)&(4.141)&(2.931)&(2.912)&(2.889)&(-0.289)&(-0.290)&(-0.288)&(2.585)&(2.671)&(2.302)\\
Year FE&Yes&Yes&Yes&Yes&Yes&Yes&Yes&Yes&Yes&Yes&Yes&Yes\\
Industry FE &Yes&Yes&Yes&Yes&Yes&Yes&Yes&Yes&Yes&Yes&Yes&Yes\\
Observations &7287&7287&7287&7287&7287&7287&7287&7287&7287&7287&7287&7287\\
$R$-squared&0.4110&0.4110&0.4110&0.3250&0.324&0.3250&0.4180&0.4180&0.4170&0.5200&0.5190&0.5200\\
\bottomrule
\end{tabular}
\smallskip
\\
Note: The dependent variable, \emph{Delay}, is calculated from Eq.~(\ref{Eq:PriceDelay}) and Eq.~(\ref{Eq:Delay}). The centrality measures include \emph{Degree}, \emph{Eigen}, and \emph{Between}. Control variables include the log of size at the end of the previous fiscal year (\emph{Size}) and book-to market ratio at the end of the previous fiscal year (\emph{Bm}), contemporaneous return on equity (\emph{Roe}), leverage at the end of the previous fiscal year (\emph{Lev}), and indicator variables for year and the industry classification. All variables are standardized to have a mean of zero and a standard deviation of one. Numbers in parentheses represent t-values that are adjusted using standard errors corrected for clustering at the firm level. *, **, *** indicate significance at the 10$\%$, 5$\%$, and 1$\%$ level, respectively.
\label{TB:MultiVarAnal:PriceDelay}
\end{table}

Table~\ref{TB:MultiVarAnal:PriceDelay} shows the results. The coefficients corresponding to the centrality measures, \emph{Degree}, \emph{Eigen} and \emph{Between}, are negative and insignificant in all the regression specifications except for those in the large-sized group (-0.0678, $t=-2.941$; -0.0653, $t=-2.612$; -0.0451, $t=-2.380$, respectively), showing a similar pattern as in Table \ref{TB:MultiVarAnal:SizeSorted} and Table~\ref{TB:MultiVarAnal:NoiseTrading}. For large firms at more central positions in the cross-shareholding network, their stocks tend to be priced more rationally and therefore reflect more contemporaneous market-wide information. Not surprisingly, consistent with our interpretation in previous analyses, large firms are more likely to have a mature and developed corporate structure and a stable future cash flow; therefore individuals prefer to invest on a long-term horizon, strengthening the alignment effect between shareholders and demanding a higher quality of the information environment. Therefore, large firms' stock would contain more contemporary information instead of historical market-wide information. Besides, in line with \cite{Pareek-2012}, firm size and leverage have negative and positive effects on the price delay, respectively.

Another potential explanation for our results may rest on the investor attention, as \cite{Pareek-2012} attributes such a price delay to the potential probability of investor attention when studying fund managers' trading behavior. Firms that are small, have high levels of arbitrage risk and transaction costs, low levels of institutional holdings, and that are largely not followed by analysts have been proved to be the impediments to informed trading on the stock market \citep{Zhang-2006-JF,Zhang-Cai-Keasey-2013-JAccE,Michaely-Rubin-Vedrashko-2016-JFE}. \cite{Cohen-Frazzini-2008-JF} conjecture that investors who buy and hold stocks whose subsidiaries recently experienced a profit and sell stocks whose subsidiaries recently suffered from a loss would gain abnormal returns. More evidence of the ownership connection, given by \cite{Ginglinger-Hebert-Renneboog-2011}, shows that the investors of the parent firm would pay more attention to the earning news of its subsidiaries, while there exists a persistent inattention to the earning news of the parent firm among investors of its subsidiaries. Therefore, since firms' balance sheets are a type of open data available to investors, we believe that investors would pay attention to the stock of firms' cross-shareholders.

On one hand, investor attention would have a direct impact on price informativeness even if investors pay attention to cross-shareholders through balance sheets, because in a financial market filled with inexperienced investors, stock pricing would be more likely to be biased or sentiment-driven when they receive more attention from other firms (e.g., herding) \citep{Engelberg-Gao-2011-JF}. In this way, among large firms, we might observe a positive (negative) or insignificant relation between noise trading (price delay) and the centrality in the cross-shareholding network. However, our results show that noise trading is negatively associated with firms' centrality in the cross-shareholding network, excluding a potential concern that firms' centrality in the network reduces the price delay only by attracting more attention from investors.

On the other hand, investor attention might have an indirect impact on pricing efficiency. Stocks would be much less likely to be mispriced when they are focused especially for firms with a good information environment and less noise traders. Therefore, in spite of a descriptive interpretation, we consider that when looking for which stocks worth holding, investors would not only think over the fundamental information of the stock they already hold, but also consider the stocks of their cross-shareholders, especially for large firms at more central positions in the cross-shareholding network that would obtain more attention than small firms. Based on the improvement of the information environment arising from the cross-shareholding for large firms, such a increasing level of attention would further strengthen the impact of the reduction of noise trading on price informativeness. Therefore, for large firms, centrality in the cross-shareholding network is positively associated with pricing efficiency, thus reducing price delay. Moreover, this negative relation between centrality and price delay would also be strengthened by more attention from investors at a more central position in the network.

Our results remain unchanged using the price delay calculated from the value-weighted industry portfolio, or using \emph{Rd}, \emph{Re} and \emph{Rb} as centrality measures\footnote{Following \cite{Pareek-2012} and \cite{Gu-Kang-Xu-2018-JBF}, we include the yearly turnover ratio of the tradable shares for the present year (\emph{Turnover}), obtained from the RESSET Financial Database, as control variables to measure the liquidity in unreported tests; our unreported results remain unaffected.}. In summary, these results prove the positive impact of cross-shareholding on stock pricing efficiency, which further promotes stock price synchronicity, especially for large firms.

\subsection{The signs of market return}

As we can see from the previous analyses, cross-shareholding improves stock price informativeness by reducing noise trading. It should come as no surprise that this effect appears to be stable if the negative impact of noise trading on stock price informativeness is maintained in the long run. However, the market environment changes from time to time. \cite{Lo-2004-JPM} sheds a light on the changing market environment, and suggests that the risk/reward relation in the stock market always varies. Based on the Adaptive Markets Hypothesis of \cite{Lo-2004-JPM}, \cite{Shi-Zhou-2017a-PA} and \cite{Shi-Zhou-2017b-PA} confirm that trading strategies constructed based on contrarian effects would fail in Chinese stock market under certain market conditions. Considering that the impact of noise trading would differ from one certain market condition to another, we next investigate the impact of cross-shareholding on stock prices in separate market environment.

Following \cite{Bris-Goetzmann-William-Zhu-2010-JF}, we calculate two separate measures of individual stock price co-movement. Let $r_m^+$ be the value-weighted market return when it is either positive or zero and $r_m^-$ be the market return when it is negative. For each stock in our sample and for each year \emph{T}, we calculate the ${R^2}$ values in the following two modified specifications:
\begin{equation}
 {r_{j,t}} = \alpha_j^+  + {\beta_j}r_{m,t}^+  + \varepsilon_{j,t}^+
 \label{Eq:Sign:pos}
\end{equation}
and
\begin{equation}
 {r_{j,t}} = \alpha_j^-  + {\beta_j}r_{m,t}^-  + \varepsilon_{j,t}^-.
 \label{Eq:Sign:neg}
\end{equation}
We compute the corresponding ${R^2}$ coefficients, $R_j^{2 + }$ and $R_j^{2 - }$, respectively, using weekly stock return data for every week \emph{t} in year \emph{T}. We then apply a logistic transformation to obtain near normally distributed and continuous variables, \emph{SYNP} and \emph{SYNN}, respectively. We also include the log of size at the end of the previous fiscal year (\emph{Size}), the book-to-market ratio at the end of the previous fiscal year (\emph{Bm}), leverage at the end of the previous fiscal year (\emph{Lev}), and contemporaneous return on equity (\emph{Roe}) as additional controls for firm characteristics. Following \cite{Hutton-Marcus-Tehranian-2009-JFE}, we also include the skewness and the kurtosis calculated separately from Eq.~(\ref{Eq:Sign:pos}) and Eq.~(\ref{Eq:Sign:neg}). All variables are standardized to have zero mean and unit standard deviation. Moreover, to better investigate the relation between cross-shareholding and $R_j^{2 - }$, we exclude the firm-years in the sample during 2008, 2015, and 2016 to avoid the potential bias caused by the financial crisis and the stock market crash.

%

We sort our sample into size terciles based on size at the end of the previous fiscal year and similar statistics are shown across different firm size. The data show that the overall level of ${R^{2-}}$ is higher than that of ${R^{2+}}$. The results indicate that there exists an asymmetric synchronicity between firm-specific return and market return. A similar pattern can be also found across different firm sizes. Furthermore, compared to large firms, small firms may be more sensitive to changes in market conditions because of their low levels of operation and management and their high levels of uncertainty in cash flow.

Table~\ref{TB:MultiVarAnal:Signs:Pos} presents the results of the relation between cross-shareholding and \emph{SYNP}. The coefficients of \emph{Degree}, \emph{Eigen}, and \emph{Between} are insignificant in the entire sample (-0.0001, $t=-0.012$; 0.0071, $t=0.714$; and -0.0010, $t=-0.094$, respectively). The coefficients of \emph{Bm} and \emph{Skew} are significantly positive and the coefficients of \emph{Roe} are insignificant. A similar pattern can be found in the small-size and  medium-size groups. In the large-size group, the coefficients of \emph{Degree}, \emph{Eigen} and \emph{Between} are positive and significant at the 10$\%$ level (0.0327, $t=1.739$; 0.0294, $t=1.726$; and 0.0342, $t=1.925$, respectively). Our results remain unchanged using $Rd$, $Re$ and $Rb$ as the centrality measures. The results show that cross-shareholding would slightly improve stock price synchronicity during market upturns, especially for large firms.

\begin{table}[h]
\caption{Centrality in the cross-shareholding network and {\emph{SYNP}}.}
\smallskip
\scriptsize
 \begin{tabular}{lcccccccccccc}
  \toprule
 &\multicolumn{3}{c} {All sample}& \multicolumn{3}{c} {Small size}&\multicolumn{3}{c} {Medium size}&\multicolumn{3}{c} {Large size}\\
\hline
{\emph{Degree}}&-0.0001& & &-0.0071& & &-0.0060& & &0.0327*& & \\
      &(-0.012)& & &(-0.370)& & &(-0.402)& & &(1.739)& & \\
{\emph{Eigen}}&&0.0071 & &&0.0060 & &&0.0007& &&0.0294* &\\
   &&(0.714) & &&(0.373) & &&(0.047)& &&(1.762)&\\
{\emph{Between}}&& & -0.0001&& &-0.0048&& &-0.0201 && &0.0342*\\
&& & (-0.094)&& &(-0.254) && &(-1.362) && &(1.925)\\
{\emph{Size}}&0.1200***&0.1190***&0.1210***&0.0840**&0.0831**&0.0839**&0.0637&0.0636&0.0642&0.0252&0.0318&0.0254\\
&(8.366)&(8.329)&(8.396)&(2.427)&(2.399)&(2.422)&(1.586)&(1.585)&(1.600)&(0.902)&(1.187)&(0.913)\\
{\emph{Bm}}&0.1370***&0.1360***&0.1370***&0.0766***&0.0759***&0.0764***&0.1560***&0.1560***&0.1570***&0.1530***&0.1550***&0.1540***\\
&(9.702)&(9.703)&(9.717)&(3.890)&(3.875)&(3.899)&(6.649)&(6.630)&(6.682)&(5.341)&(5.403)&(5.372)\\
{\emph{Skew}}&0.0252**&0.0250**&0.0252**&-0.0228&-0.0233&-0.0229&0.0222&0.0221&0.0224&0.0563**&0.0557**&0.0566**\\
&(2.126)&(2.103)&(2.127)&(-1.158)&(-1.187)&(-1.164)&(1.087)&(1.077)&(1.097)&(2.48)&(2.449)&(2.491)\\
{\emph{Kurt}}&-0.0939***&-0.0939***&-0.0939***&-0.0720***&-0.0720***&-0.0721***&-0.1020***&-0.1020***&-0.1020***&-0.1160***&-0.1150***&-0.1150***\\
&(-7.658)&(-7.655)&(-7.659)&(-3.748)&(-3.744)&(-3.748)&(-5.111)&(-5.108)&(-5.107)&(-4.552)&(-4.494)&(-4.538)\\
{\emph{Roe}}&0.0006&0.0006&0.0006&-0.0011&-0.0012&-0.0011&-0.0002&-0.0002&-0.0004&0.0035&0.0035&0.0031\\
&(0.087)&(0.089)&(0.089)&(-0.073)&(-0.078)&(-0.076)&(-0.009)&(-0.012)&(-0.019)&(0.503)&(0.509)&(0.419)\\
{\emph{Lev}}&-0.0155&-0.0152&-0.0156&-0.0047&-0.0047&-0.0047&-0.0190&-0.0187&-0.0195&-0.0166&-0.0181&-0.0175\\
&(-1.293)&(-1.268)&(-1.297)&(-0.293)&(-0.293)&(-0.288)&(-0.842)&(-0.831)&(-0.865)&(-0.552)&(-0.602)&(-0.581)\\
{\emph{Constant}}&-0.0657&-0.0690&-0.0652&-0.1830*&-0.1780*&-0.1810*&0.7170**&0.7170**&0.7190**&-0.0227&-0.0302&-0.0201\\
&(-0.964)&(-1.014)&(-0.956)&(-1.740)&(-1.702)&(-1.727)&(2.290)&(2.300)&(2.276)&(-0.215)&(-0.285)&(-0.195)\\
Year FE&Yes&Yes&Yes&Yes&Yes&Yes&Yes&Yes&Yes&Yes&Yes&Yes\\
Industry FE &Yes&Yes&Yes&Yes&Yes&Yes&Yes&Yes&Yes&Yes&Yes&Yes\\
Observations &7287&7287&7287&2425&2425&2425&2432&2432&2432&2430&2430&2430\\
$R$-squared&0.2870&0.2870&0.2870&0.3360&0.3360&0.3360&0.3270&0.3270&0.3270&0.2250&0.2250&0.2250\\
\bottomrule
\end{tabular}
\smallskip
\\
Note: The dependent variable, \emph{SYNP}, is calculated from Eq.~(\ref{Eq:Sign:pos}). The centrality measures include \emph{Degree}, \emph{Eigen} and \emph{Between}. Control variables include the log of size at the end of the previous fiscal year (\emph{Size}) and book-to market ratio at the end of the previous fiscal year (\emph{Bm}), skewness (\emph{Skew}), kurtosis (\emph{Kurt}), contemporaneous return on equity (\emph{Roe}), leverage at the end of the previous fiscal year (\emph{Lev}), and indicator variables for year and the industry classification. All variables are standardized to have a mean of zero and a standard deviation of one. Numbers in parentheses represent t-values that are adjusted using standard errors corrected for clustering at the firm level. *, **, *** indicate significance at the 10$\%$, 5$\%$, and 1$\%$ level, respectively.
\label{TB:MultiVarAnal:Signs:Pos}
\end{table}

Table~\ref{TB:MultiVarAnal:Signs:Neg} presents the results of the relation between cross-shareholding and \emph{SYNN}. The coefficients of \emph{Degree}, \emph{Eigen} and \emph{Between} are all significant at the 1\% level in the entire sample (0.0431, $t=3.050$; 0.0377, $t=2.926$; and 0.0372, $t=2.781$, respectively). Moreover, skewness is negatively associated with \emph{SYNN} and the coefficients of \emph{Roe} and \emph{Lev} are insignificant. A similar pattern can be found in the small-size and medium-size groups. In the large-size group, the coefficients of \emph{Degree}, \emph{Eigen} and \emph{Between} are also positive and significant at the 1\% level (0.1010, $t=4.331$; 0.0908, $t=4.676$; and 0.0932, $t=4.354$, respectively). Our results remain unchanged using $Rd$, $Re$ and $Rb$ as the centrality measures\footnote{All the results from tests, mentioned in the previous analyses where we use $Rd$, $Re$ and $Rb$ as the centrality measures to investigate the impact of cross-shareholding, are available upon request.}. The results show that cross-shareholding would strongly improve the stock price synchronicity during market downturns, especially for large firms\footnote{We re-run our regression using the firm-year sample for the period of 2004-2016 and test the firm-years using daily stock-return data. Our unreported results remain unaffected.}.

\begin{table}[h]
\caption{Centrality in the cross-shareholding network and {\emph{SYNN}}.}
\smallskip
\scriptsize
 \begin{tabular}{lcccccccccccc}
  \toprule
 &\multicolumn{3}{c} {All sample}& \multicolumn{3}{c}{Small size}&\multicolumn{3}{c}{Medium size}&\multicolumn{3}{c}{Large size}\\
\hline
{\emph{Degree}}&0.0431***& & &0.0109& & &0.0018& & &0.1010***& & \\
      &(3.050)& & &(0.528)& & &(0.072)& & &(4.331)& & \\
{\emph{Eigen}}&&0.0377*** & &&-0.0080 & &&0.0244& &&0.0908***&\\
   &&(2.926) & &&(-0.377)& &&(1.255)& &&(4.68)&\\
{\emph{Between}}&& & 0.0372***&& &0.0204&& &-0.0179&& &0.0932***\\
&& & (2.781)&& &(0.927) && &(-0.571) && &(4.354)\\
{\emph{Size}}&-0.1290***&-0.1220***&-0.1270***&0.0046&0.0056&0.0043&-0.0034&-0.0039&-0.0030&-0.1520***&-0.1310***&-0.1470***\\
&(-7.233)&(-7.072)&(-7.104)&(0.129)&(0.157)&(0.121)&(-0.084)&(-0.095)&(-0.074)&(-5.200)&(-4.744)&(-5.023)\\
{\emph{Bm}}&0.1100***&0.1110***&0.1110***&0.0340&0.0351&0.0339&0.1080***&0.1070***&0.1100***&0.1580***&0.1610***&0.1600***\\
&(6.010)&(6.041)&(6.054)&(1.168)&(1.196)&(1.165)&(3.745)&(3.699)&(3.789)&(4.851)&(4.923)&(4.924)\\
{\emph{Skew}}&-0.1160***&-0.1150***&-0.1160***&-0.1530***&-0.1530***&-0.1530***&-0.0925***&-0.0919***&-0.0925***&-0.0846***&-0.0838***&-0.0850***\\
&(-8.116)&(-8.073)&(-8.138)&(-5.697)&(-5.695)&(-5.685)&(-4.070)&(-4.045)&(-4.062)&(-3.221)&(-3.135)&(-3.265)\\
{\emph{Kurt}}&-0.0410***&-0.0415***&-0.0409***&-0.0465&-0.0463&-0.0463&-0.0507**&-0.0500**&-0.0513**&-0.0464*&-0.0470*&-0.0448*\\
&(-2.797)&(-2.826)&(-2.795)&(-1.561)&(-1.553)&(-1.554)&(-2.164)&(-2.136)&(-2.181)&(-1.928)&(-1.916)&(-1.871)\\
{\emph{Roe}}&-0.0052&-0.0051&-0.0055&0.0136&0.0139&0.0136&-0.0285&-0.0290&-0.0285&-0.0069&-0.0069&-0.0083\\
&(-0.597)&(-0.584)&(-0.658)&(0.352)&(0.357)&(0.350)&(-1.185)&(-1.201)&(-1.182)&(-0.685)&(-0.671)&(-0.908)\\
{\emph{Lev}}&-0.0284*&-0.0295*&-0.0292*&0.0013&0.0011&0.0013&-0.0445&-0.0437&-0.0452&-0.0426&-0.0462&-0.0466\\
&(-1.662)&(-1.713)&(-1.704)&(0.054)&(0.049)&(0.055)&(-1.450)&(-1.427)&(-1.481)&(-1.286)&(-1.384)&(-1.392)\\
{\emph{Constant}}&-1.6930***&-1.6850***&-1.6930***&0.0826&0.0740&0.0810&0.4050**&0.4020*&0.4150**&-1.1490***&-1.1420***&-1.1650***\\
&(-22.270)&(-22.260)&(-22.200)&(0.267)&(0.238)&(0.261)&(1.986)&(1.961)&(2.032)&(-2.878)&(-2.832)&(-2.909)\\
Year FE&Yes&Yes&Yes&Yes&Yes&Yes&Yes&Yes&Yes&Yes&Yes&Yes\\
Industry FE &Yes&Yes&Yes&Yes&Yes&Yes&Yes&Yes&Yes&Yes&Yes&Yes\\
Observations &5569&5569&5569&1853&1853&1853&1859&1859&1859&1857&1857&1857\\
$R$-squared&0.0910&0.0900&0.0900&0.1230&0.1230&0.1230&0.0950&0.0960&0.0950&0.1200&0.1190&0.1190\\
\bottomrule
\end{tabular}
\smallskip
\\
Note: The dependent variable, \emph{SYNN}, is calculated from Eq.~(\ref{Eq:Sign:neg}). The centrality measures include \emph{Degree}, \emph{Eigen}, and \emph{Between}. Control variables include the log of size at the end of the previous fiscal year (\emph{Mv}) and book-to market ratio at the end of the previous fiscal year (\emph{Bm}), skewness (\emph{Skew}), kurtosis (\emph{Kurt}), contemporaneous return on equity (\emph{Roe}), leverage at the end of the previous fiscal year (\emph{Lev}), and indicator variables for year and the industry classification. All variables are standardized to have a mean of zero and a standard deviation of one. Numbers in parentheses represent t-values that are adjusted using standard errors corrected for clustering at the firm level. *, **, *** indicate significance at the 10$\%$, 5$\%$, and 1$\%$ level, respectively.
\label{TB:MultiVarAnal:Signs:Neg}
\end{table}

\interfootnotelinepenalty=10000

When we turn to compare the results for large firms in Table~\ref{TB:MultiVarAnal:Signs:Pos} and Table~\ref{TB:MultiVarAnal:Signs:Neg}, the coefficients of \emph{Degree} \emph{Eigen}, and \emph{Between} during market downturns are higher and more significant. Moreover, compared to the average levels of the coefficients of the centrality of large firms in Table~\ref{TB:MultiVarAnal:SizeSorted}, the coefficients of \emph{Degree}, \emph{Eigen}, and \emph{Between} are lower in Table~\ref{TB:MultiVarAnal:Signs:Pos} and higher in Table~\ref{TB:MultiVarAnal:Signs:Neg}. It is remarkable that the impact of cross-shareholding on stock price informativeness has an asymmetric pattern, implying that stock prices become more synchronous with market returns during market downturns. Consistent with \cite{Lo-2004-JPM}'s conjecture, the results show that the impact of cross-shareholding on price informativeness are highly dependent on the market environment as the impact of noise trading on pricing efficiency varies in the Chinese stock market.

More concretely, in the financial market, investors' behavior matters since it is also a type of dynamic decision-making behavior \citep{Wen-He-Chen-2014-MPE,Wen-He-Gong-Liu-2014-DDNS}. \cite{Hou-Xiong-Peng-2009} find that underreaction among investors would occur during market upturns and overreaction would occur during market downturns, while \cite{Frank-Sanati-2018-JFE} argue that the stock market overreacts to good news and underreacts to bad news. As mentioned in the analyses of price delay, we argue that investor attention not only has a direct impact but also is likely to make a difference on the impact of reduction of noise trading on price informativeness. Therefore, in spite of a descriptive interpretation, we reckon that, in the Chinese stock market, investors would pay more attention to the portion of firms' information that is related to market-wide information during market downturns with a potential goal of bargain-hunting or stop-loss strategies. Therefore, investors accelerate the incorporation of market-wide information into the stock price, and thus strengthen the impact of the reduction in noise trading on the process of price discovery during market downturns.

To some extent, short selling may cast doubt on our results. While prices incorporate negative information faster in markets where short selling is allowed and practiced \citep{Bris-Goetzmann-William-Zhu-2010-JF}, we do not think that it looks like the fact that such an effect is derived from the removal of short selling. In March 2010, the CSRC launched a plan to remove the ban on short selling; we carry out a complementary analysis using the firm-years in the sample in 2004-2009 to avoid potential selection bias. Our inferences remain unaffected in these unreported tests. In addition, one can find that the effect stems from those firms with limits on short selling. Given that high arbitrage limits increase the possibility and magnitude of the potential for market mispricing \citep{Gu-Kang-Xu-2018-JBF}, we infer that firms at more central positions in the cross-shareholding network have a higher level of noise trading, which is not consistent with our conjecture in the previous analyses. Therefore, such a concern does not make sense.

In summary, the impact of cross-shareholding on stock price informativeness is highly dependent on the market environment, instead of being isolated from the outside world and remaining static. Specifically, this impact becomes stronger during market downturns. Our results provide further evidence supporting the Adaptive Markets Hypothesis of \cite{Lo-2004-JPM} that market efficiency varies from time to time.

\section{Conclusions}
\label{S1:Conclude}

In this paper, we investigate whether and how stock price synchronicity, as a measure of stock price informativeness, is associated with firms' centrality in the cross-shareholding network unique to China.
We obtain five main findings.

Contrary to the conventional wisdom that the information environment specializes in the production of firm-specific information, we find that the improvement of the information environment reduces noise trading in the Chinese stock market. First, using the ${R^2}$ of an expanded model as a measure of stock price synchronicity, we find that a higher centrality of a firm in the cross-shareholding network increases the stock price co-movement with the market and the industry portfolio return. Second, to further confirm the positive association between stock price synchronicity and the information environment \citep{Lee-Liu-2011-JBF}, as a result of cross-shareholding, we confirm that it is through a noise-reducing process that cross-shareholding promotes stock price synchronicity in the Chinese stock market. Third, according to the noise-reducing process and the potential for investor inattention, we find that firms with a more central position in the cross-shareholding network have a lower level of stock price delay. Fourth, the effects of cross-shareholding are pronounced for large firms. Apart from the phenomenon that noise trading is clustered more at small firms, we suggest that large firms have more stable future cash flows and lower sensitivity to market-wide shocks, which strengthens the alignment effect between their shareholders and demands a higher quality of the information environment. Lastly, cross-shareholding have asymmetric impacts on stock price synchronicity between market upturns and market downturns. We suggest that investors in the Chinese market would pay more attention to firms' information during market downturns, thus strengthening the impact of the reduction in noise trading on the process of price discovery.

Our findings have some implications for the existing literature. First, our work is related to \cite{Khanna-Thomas-2009-JFE} who find that the pairwise stock price synchronicity is more strongly correlated with interlocking directorates, while there is a correlation between synchronicity and equity ties. Considering the equity linkage, we employ complex network theory to analyze the relation between cross-shareholding and stock price synchronicity. Second, while researchers have focused on how the ownership structure affects the information environment of a firm and its decision making, we test further the impacts of cross-shareholding on the variations of stock prices. Third, we extend the theoretical predictions \citep{Lee-Liu-2011-JBF} by providing empirical evidence which shows an inverse relation between noise trading and stock price synchronicity.

The results also suggest avenues for future work. It would be useful to distinguish the financial interlocks from different directions and weights so as to investigate whether cross-shareholders, who are also large controlling shareholders, have an impact on stock price informativeness. In addition, we do not take a comprehensive approach towards identifying the mechanisms through whether and how investor inattention varies across firms with different levels of cross-shareholding, arguing only that investors may pay attention to cross-shareholders through the balance sheets of firms. An open question is whether and how investor inattention varies across firms in a financial network \citep{Richardson-Tuna-Wysocki-2010-JAE}.

\section*{Acknowledgment}

We gratefully acknowledge the financial support from the National Natural Science Foundation of China (nos. 71873146, 71873147, 71431008, U1811462).



\end{document}